\documentclass[11pt]{article}
\usepackage{fullpage,mathptm,times,amsmath,mathdots,amssymb,epsf,graphicx}
\advance\parskip 4pt

\newtheorem{theorem}{Theorem}[section]
\newtheorem{definition}[theorem]{Definition}
\newtheorem{lemma}[theorem]{Lemma}

\newtheorem{corollary}[theorem]{Corollary}

\newenvironment{example}{\refstepcounter{theorem}
  \par\kern\medskipamount\noindent\textit{Example~\thetheorem}~}{\par\smallskip}
\def\proof{\par\kern-\medskipamount\noindent\textbf{Proof.}~~}
\def\remark{\par\kern\medskipamount\noindent\textbf{Remark.}~}
\def\nremark{\par\kern\medskipamount\noindent\textbf{Remark~\stepcounter{theorem}\arabic{section}.\arabic{theorem}}~}

\def\theequation{\arabic{section}.\arabic{equation}}
\catcode`\@ 11
\@addtoreset{equation}{section}
\def\overl@ss#1#2{\vcenter{\offinterlineskip
        \ialign{$\m@th#1\hfil##\hfil$\crcr#2\crcr\raise0.6ex\hbox{$<$}\crcr } }}
\def\overgr@ater#1#2{\vcenter{\offinterlineskip
        \ialign{$\m@th#1\hfil##\hfil$\crcr#2\crcr\raise0.6ex\hbox{$>$}\crcr } }}
\def\gl{\mathrel{\mathpalette\overl@ss>}}
\def\lg{\mathrel{\mathpalette\overgr@ater<}}
\long\def\@makecaption#1#2{\vskip\abovecaptionskip
  \sbox\@tempboxa{\small #1: #2}%
  \ifdim \wd\@tempboxa >\hsize \small #1: #2\par
  \else \global \@minipagefalse \hb@xt@\hsize{\hfil\box\@tempboxa\hfil}\fi
  \vskip\belowcaptionskip}
\catcode`\@ 12

\let\_=\mathsf
\let\==\bar
\let\@=\mathbf
\newcommand{\partialderiv}[3][]{\frac{\partial^{#1}#2}{\partial {#3}^{#1}}}
\def\circ{\ifmmode\mathchar"220E\else$\mathchar"220E$\fi}
\def\Wr{\mathop{\mathrm{Wr}}\nolimits}
\let\trueint=\int
\let\truesum=\sum
\def\int{\mathop{\textstyle\trueint}\limits}
\def\sum{\mathop{\textstyle\truesum}\limits}
\def\diag{\mathop{\rm diag}\nolimits}
\let\<=\langle
\let\>=\rangle
\def\Real{{\mathbb{R}}}
\def\sech{\mathop{\rm sech}\nolimits}

\def\half{{\textstyle\frac12}}
\def\rank{\mathop{\rm rank}\nolimits}
\let\next=\phi\global\let\phi=\varphi\global\let\varphi=\next
\renewcommand\labelitemi{\ifmmode\circ\else$\circ$\fi}
\allowdisplaybreaks
\def\A{{\mathcal A}}
\def\e{{\mathrm e}}
\def\mapto#1#2{\mathop{\longrightarrow}\limits^{#1\to#2}}
\def\note#1{\relax}
\def\today{October 31, 2005}

\begin{document}
\title{\bf Soliton solutions of the Kadomtsev-Petviashvili II equation}
\author{Gino Biondini$^1$ and Sarbarish Chakravarty$^2$\\[1ex]
\small$^1$\it\ 
State University of New York, Department of Mathematics, Buffalo, NY 14260-2900\\
\small$^2$\it\
University of Colorado, Department of Mathematics, Colorado Springs, CO 80933-7150}
\date{\small\today}
\maketitle
\begin{abstract}
\noindent
We study a general class of line-soliton solutions of the
Kadomtsev-Petviashvili II (KPII) equation
by investigating the Wronskian form of its tau-function.
We show that, in addition to previously known line-soliton solutions, 
this class also contains a large variety of new multi-soliton solutions,
many of which exhibit nontrivial
spatial interaction patterns. 
We also show that, in general, such solutions consist of 
unequal numbers of incoming and outgoing line solitons.
From the asymptotic analysis of the tau-function,
we explicitly characterize the incoming and outgoing
line-solitons of this class of solutions. 
We illustrate these results by discussing several examples.
\end{abstract}

\section{Introduction}
\label{s:introduction}

The Kadomtsev-Petviashvili (KP) equation
\begin{equation}
\partialderiv{}x\left(-4\partialderiv ut
  +\partialderiv[3]ux +6u\partialderiv ux \right)
            + 3\sigma^2 \partialderiv[2]uy =0\,
\label{e:KP}
\end{equation}
where $u=u(x,y,t)$ and $\sigma^2=\pm1$,
is one of the prototypical (2+1)-dimensional integrable nonlinear
partial differential equations. The case $\sigma^2=-1$ is known as the KPI equation,
and $\sigma^2=1$ as the KPII equation. Originally derived \cite{SovPhysDoklady15p539}
as a model for small-amplitude, long-wavelength, weakly two-dimensional (y-variation much slower
than the x-variation) solitary waves in a weakly dispersive medium,
the KP equation arises in disparate physical settings including 
water waves and plasmas, astrophysics, cosmology, optics, magnetics,
anisotropic two-dimensional lattices and
Bose-Einstein condensation.
The remarkably rich mathematical structure underlying the KP equation,
its integrability and large classes of exact solutions have been studied
extensively for the past thirty years, and are documented in several monographs
\unskip~\cite{AblowitzClarkson,Dickey,InfeldRowlands,MatveevSalle,MiwaJimboDate,NMPZ1984}.

In this article we study a large class of solitary wave solutions
of the KPII equation. 
It is well-known (e.g., see Refs.~\cite{PLA95p1,MatveevSalle})
that solutions of the KPII equation can be expressed as
\begin{equation}
u(x,y,t)= 2\partialderiv[2]{ }x\log\tau(x,y,t)\,,
\label{e:u}
\end{equation}
where the tau~function $\tau(x,y,t)$ is given in terms of the
Wronskian determinant~\cite{Hirota,MatveevSalle}
\begin{equation}
\tau(x,y,t)= \Wr(f_1,\dots,f_N)= 
  \begin{pmatrix}
     f_1 & f_2 & \cdots & f_N \\
     f_1' & f_2' &\cdots & f_N' \\
     \vdots & \vdots & & \vdots \\
     f_N^{(N-1)} & f_2^{(N-1)} & \cdots &f_N^{(N-1)}
  \end{pmatrix}\,.
\label{e:tau}
\end{equation}
with $f^{(i)}= \partial^i \!f/\partial x^i$, 
and where the functions $f_1,\dots,f_N$ 
are a set of linearly independent solutions of the linear system
\begin{equation}
\partialderiv fy= \partialderiv[2]fx\,,
\qquad
\partialderiv ft= \partialderiv[3]fx\,.
\label{e:u0Laxpair}
\end{equation}
It should be noted that Eq.~\eqref{e:tau} can also be obtained
as the composition of $N$~Darboux transformations for KPII~\cite{MatveevSalle}.
Note also that the Lax pair of the KP equation is given by~\cite{Dryuma}
$\sigma \partial_y f - \partial_x^2 f + u f = 0$
and 
$\partial_t f - \partial_x^3 f + 6u(\partial_x u) f
 + 3\sigma (\partial_x^{-1}\partial_y u) f = 0$.
Thus, the functions $f_1,\dots,f_N$ in Eqs.~\eqref{e:tau} 
are precisely $N$ solutions of the zero-potential Lax pair of~KPII.
A one-soliton solution of the KPII equation
is obtained by choosing $N=1$ and
$f(x,y,t)= e^{\theta_1}+e^{\theta_2}$,
where 
\begin{equation}
\theta_m(x,y,t) = k_mx +k_m^2y+k_m^3t+\theta_{m,0}
\label{e:theta}
\end{equation}
with $k_m,\theta_{m,0}\in\Real,~m=1,2$ and with $k_1 \ne k_2$ 
for nontrivial solutions. 
Without loss of generality, one can order the parameters as $k_1 < k_2$. 
The above choice yields the following traveling-wave solution 
\begin{equation}
u(x,y,t)= \half(k_2-k_1)^2\sech^2\half(\theta_2-\theta_1)= \Phi(\@k\cdot\@x+\omega t) \,, 
\label{e:onesoliton}
\end{equation}
where $\@x=(x,y)$.
The wavevector $\@k=(l_x,l_y)$ and the frequency $\omega$
are given by
\begin{equation}
\@k=(k_1-k_2,k_1^2-k_2^2)\,,\qquad
\omega=k_1^3-k_2^3\,,
\end{equation}
and they satisfy the nonlinear dispersion relation
\begin{equation}
-4\omega l_x+l_x^4+3l_y^2=0\,.
\label{e:dispersionrelation}
\end{equation}
The solution in Eq.~\eqref{e:onesoliton} is localized along points
satisfying the equation $\theta_1=\theta_2$,
which defines a line in the the $xy$-plane,
Such solitary wave solutions of the KPII equation are thus called 
\textit{line solitons}.
They are stable with respect to transverse perturbations. 
It is worth mentioning here that the KPI equation 
(namely, Eq.~\eqref{e:KP} with $\sigma^2=-1$) 
also admits line-soliton solutions, but these solutions 
are not stable with respect to small tranverse perturbations.

Equation~\eqref{e:onesoliton} also implies that, 
apart from a constant $\theta_{1,0}-\theta_{2,0}$ 
(corresponding to an overall translation of the solution),
a line soliton of KPII is characterized by either the phase parameters
$k_1, k_2$, or by two physical parameters, namely,
the \textit{soliton amplitude}~$a$ 
and the \textit{soliton direction}~$c$, defined respectively as
\begin{equation}
a = k_2-k_1\,,\qquad
c= k_1+k_2\,.
\end{equation}
Note that $c=\tan \alpha$, 
where $\alpha$ is the angle, measured counterclockwise,
between the line soliton and the positive $y$-axis.
Hence, the soliton direction~$c$ can also be viewed as the ``velocity'' of 
the soliton in the $xy$-plane: $c= -dx/dy = l_y/l_x$.
For any given choice of amplitude and direction of the soliton,
one obtains the phase parameters $k_{1,2}$ uniquely as
$k_1=\half (c-a)$ and $k_2=\half (c+a)$. 
Finally, note that when $c=0$ (equivalently, $k_1 = -k_2$), the solution 
in Eq.~\eqref{e:onesoliton} becomes $y$-independent and reduces
to the one-soliton solution of the Korteweg-de~Vries (KdV) equation.

Similar to KdV, it is also possible to obtain multi-soliton solutions of
the KPII equation.
As $y\to\pm\infty$, each of these multi-soliton solutions
consists of a number of line solitons which are exponentially separated,
and are sorted according to their directions, with
increasing values of~$c$ from left to right as $y\to-\infty$
and increasing values of~$c$ from right to left as $y\to\infty$.
However, the multi-soliton solution space of the KPII equation
turns out to be much richer than that of the (1+1)-dimensional KdV
equation due to the dependence of the KPII solutions on
the additional spatial variable $y$.

It is possible to construct a general family of multi-soliton solutions
via the Wronskian formalism of Eq.~\eqref{e:tau}
by choosing~$M$ phases $\theta_1,\dots,\theta_M$
defined as in Eq.~\eqref{e:theta} 
with distinct real \textit{phase parameters} $k_1 < k_2 < \ldots < k_M$
and then considering 
the functions $f_1,\dots,f_N$
in Eq.~\eqref{e:tau} defined by
\begin{equation}
f_n(x,y,t)= \sum_{m=1}^{M} a_{n,m}\,e^{\theta_m}\,, \quad 
n = 1,2, \ldots, N\,,
\label{e:f}
\end{equation}
The constant coefficients $a_{n,m}$ define the $N \times M$ 
\textit{coefficient matrix}
$A:= (a_{n,m})$, which is required to be of full rank
(i.e., $\mathop{\rm rank}(A)=N$) 
and all of whose non-zero $N \times N$ minors must be sign definite.
The full rank condition is necessary and sufficient for the
functions $f_n$ in Eq.~\eqref{e:f} to be linearly independent.
The sign definiteness of the non-zero minors is sufficient to 
ensure that the tau~function $\tau(x,y,t)$
has no zeros in the $xy$-plane for all $t$, 
so that the KPII solution $u(x,y,t)$ resulting from Eq.~\eqref{e:u} is 
non-singular.

One of the main results of this work (Theorem~\ref{A:T1})
is to show that, when the coefficient matrix $A$ satisfies certain
irreducibility conditions (cf.\ Definition~\ref{A:D1}),
Eq.~\eqref{e:f} leads to a multi-soliton configuration which consists of 
$N_-$ asymptotic line solitons as $y \rightarrow -\infty$
and $N_+$ asymptotic line solitons as $y \rightarrow \infty$, 
with $N_- = M-N$ and $N_+=N$,   
and where each of the asymptotic line solitons has the form of 
a plane wave similar to the one-soliton solution in Eq.~\eqref{e:onesoliton}. 
We refer to these multi-soliton configurations as the 
\textit{$(N_-,N_+)$-soliton solutions} of KPII; also, we will call 
\textit{incoming} line solitons the asymptotic line solitons as $y\to-\infty$
and \textit{outgoing} line solitons those as $y\to\infty$.
The amplitudes, directions and even the number of incoming solitons
are in general different from those of the outgoing ones,
depending on the values of $M$, $N$, the phase parameters $k_1,\dots,k_M$
and the coefficient matrix~$A$. 
Moreover, these multi-soliton solutions of KPII
exhibit a variety of spatial interaction patterns which include
the formation of intermediate line solitons and web structures
in the $xy$-plane~\cite{jphysa36p10519,GBSCYK,Kodama,medina,pashaev}.
In contrast, for the previously 
known~\cite{JPSJ1976v40p286,PLA95p1,MatveevSalle} \textit{ordinary} 
soliton solutions of KPII 
(cf.\ section~\ref{s:examples}) and solutions of KdV
the solitons experience only a phase shift after collision.
In several cases studied so far,
the existence of these nontrivial spatial features 
was found to be related to the presence of \textit{resonant} soliton 
interactions~\cite{JFM1977v79p171,NewellRedekopp,JPSJ1983v52p749}.
Several examples of these novel $(N_-,N_+)$-soliton solutions of KPII
are discussed throughout this work
(e.g., see Figs.~\ref{f:dominantphases}--\ref{f:kps}).

If $M=2N$, it follows that $N_-=N_+=N$, 
i.e., the numbers incoming and outgoing asymptotic line solitons are the same;
we call the resulting solitons the \textit{$N$-soliton solutions} of KPII.
Among these, there is an important sub-class of solutions, 
for which the amplitudes and directions of the outgoing line solitons
coincide with those of the incoming line solitons;
we call these the \textit{elastic} $N$-soliton solutions of KPII.
Elastic $N$-soliton solutions possess a number of interesting features 
of their own,
and their specific properties are further studied 
in Refs.~\cite{GBSCYK,Kodama}.

We note that multi-soliton solutions exhibiting 
nontrivial spatial structures and interaction patterns 
were also recently found in other (2+1)-dimensional integrable equations.
For example, solutions with soliton resonance and web structure 
were presented in Refs.~\cite{jphysa35p6893,jphysa36p9533}
for a coupled KP system,
and similar solutions were also found in Ref.~\cite{jphysa37p11819} 
in discrete soliton systems such as the two-dimensional Toda lattice, 
together with its fully-discrete and ultra-discrete analogues.
In other words, the existence of these solutions appears to be 
a rather common feature of (2+1)-dimensional integrable systems.
Thus, we expect that the scope of the results described in this work 
will not be limited to the KP equation alone, but will also be applicable 
to a variety of other (2+1)-dimensional integrable systems.

\section{The tau-function and the asymptotic line solitons}
\label{s:general}

In this section we investigate the properties of the tau-function
in Eq.~\eqref{e:tau}
when the $N$ functions $f_1,\dots,f_N$ are 
are chosen according to Eq.~\eqref{e:f} 
as linear combinations of $M$ exponentials  
$\e^{\theta_1}, \ldots, \e^{\theta_M}$. 
We should emphasize that Eq.~\eqref{e:f} represents the most general form 
for the functions involving linear combinations of exponential phases.
Since the elements of the $N\times M$ coefficient matrix~$A=(a_{n,m})$
are the linear combination coefficients of the functions $f_1,\dots,f_N$, 
one can naturally identify each $f_n$ 
with one of the rows of~$A$ and each phase~$\theta_m$ with one of 
the columns of~$A$, and viceversa.
In this section we examine the asymptotic behavior of the 
tau-function in the $xy$-plane as $y\to\pm\infty$. 
It is clear that, 
with the above choice of functions, the tau-function is
a linear combination of exponentials.
Consequently, the leading order behavior of the tau-function 
as $y\to\pm\infty$ in a given asymptotic sector of the $xy$-plane 
is governed by those exponential terms which are dominant in that sector. 
A systematic analysis of the dominant exponential phases allows us 
to characterize the incoming and outgoing line solitons of 
$(N_-,N_+)$-soliton solutions of~KPII.

\subsection{Basic properties of the tau-function}

We start by presenting some general properties of the tau-function.
Without loss of generality, throughout this work we choose 
the phase parameters $k_m$ to be distinct and well-ordered as  
$k_1<k_2<\dots<k_M$.
%
\begin{lemma}
\label{L:binetcauchy}
Suppose $\tau_{N,M}=\Wr(f_1,\dots,f_N)$ as in Eq.~\eqref{e:tau},
with the functions $f_1,\dots,f_N$ given by Eq.~\eqref{e:f}.
Then
\begin{equation}
\tau_{N,M}(x,y,t)= \det(A\,\Theta\,K^T)\,,
\label{e:taudet}
\end{equation}
where $A=(a_{n,m})$ is the $N\times M$ coefficient matrix,
$\Theta= \diag(\e^{\theta_1}, \ldots, \e^{\theta_M})$,
and the $N\times M$ matrix $K$ is given by 
\[
K= \begin{pmatrix}
  1 & 1 & \cdots & 1\\
  k_1 & k_2 &\cdots & k_M\\
  \vdots  & \vdots & & \vdots\\
  k_1^{N-1} & k_2^{N-1} & \cdots & k_M^{N-1}\\
  \end{pmatrix}\,,
\]
where the superscript $T$ denotes matrix transpose.
Moreover, $\tau_{N,M}$ can be expressed as
\begin{equation}
\tau_{N,M}(x,y,t)= \sum_{1\le m_1<m_2<\dots<m_N\le M} 
  V(m_1,\dots,m_N) \,\,
  A(m_1,\dots,m_N) \,\,
  \exp[\,\, \theta_{m_1,\dots,m_N} \,]\,,
\label{e:tauphases}
\end{equation}
where $\theta_{m_1,\dots,m_N}$ denotes the phase combination
\begin{equation}
 \theta_{m_1,\dots,m_N}(x,y,t) =
 \theta_{m_1}(x,y,t)+\cdots+\theta_{m_N}(x,y,t)
  \,,
\label{e:phasecombinations}
\end{equation}
$A(m_1,\dots,m_N)$ denotes the $N\times N$ minor of $A$
obtained by selecting columns ${m_1,\dots,m_N}$,
and $V(m_1,\dots,m_N)$ denotes the Van~der~Monde determinant
\begin{equation}
  V(m_1,\dots,m_N)= \prod_{1\le s_1 < s_2\le N}(k_{m_{s_2}}-k_{m_{s_1}})\,.
\label{e:vandermonde}
\end{equation}
\end{lemma}
\begin{proof}
Equation~\eqref{e:taudet} follows 
by direct computation of the Wronskian determinant~\eqref{e:tau}.
Next, to prove Eq.~\eqref{e:tauphases}
apply the Binet-Cauchy theorem to expand the determinant
in Eq.~\eqref{e:taudet} and note that the $N\times N$ minor of~$K$
obtained by selecting columns $1\le m_1<\dots<m_N\le M$ is given by
the Van~der~Monde determinant~$V(m_1,\dots,m_N)$.~\unskip
\end{proof}

From Lemma~\ref{L:binetcauchy} we have
the following basic properties of the tau-function:
\begin{enumerate}
\item
The spatio-temporal dependence of the tau-function 
in Eq.~\eqref{e:tauphases} is confined to a sum of 
exponential phase combinations $\theta_{m_1,\dots,m_N}$ 
which according to Eq.~\eqref{e:phasecombinations} are linear in $x,y,t$.
Moreover, all the Van der Monde determinants $V(m_1,\dots,m_N)$ are positive,
as the phase parameters $k_1,\dots,k_M$ are well-ordered.
Note that a sufficient condition for the tau-function~\eqref{e:tauphases} 
to generate a non-singular solution of KPII is that it is sign-definite 
for all~$(x,y,t)\in\Real^3$.
In turn, 
a sufficient condition for the tau-function~\eqref{e:tauphases} 
to be sign-definite is that the minors of the coefficient matrix~$A$ 
are either all non-negative or all non-positive.
Note however that it is not clear at present whether 
these conditions are also necessary.
\item
Each exponential term in the tau-function of Eq.~\eqref{e:tauphases} 
contains combinations of 
$N$~\textit{distinct} phases $\theta_{m_1},\dots,\theta_{m_N}$
identified by integers $m_1,\dots,m_N$ chosen from $\{1,\dots,M\}$.
Thus, the maximum number of terms in the tau-function is given by 
the binomial coefficient~$\binom{M}{N}$.
However, a given phase combination $\theta_{m_1,\dots,m_N}$ is actually
\textit{present} in the tau-function if and only if the corresponding 
minor $A(m_1,\dots,m_N)$ is non-zero.
\item
If $M<N$ the functions $f_1,\dots,f_N$ are linearly dependent;
in this case there are no terms in the summation in Eq.~\eqref{e:tauphases},
and therefore the tau-function $\tau_{N,M}(x,y,t)$ is identically zero.
Also, if $M=N$, there is only one term in the summation 
(corresponding to the determinant of~$A$);
then $\tau_{N,M}(x,y,t)$ depends linearly on $x$ 
and therefore it generates the trivial solution of~KP.
Finally, if $\mathop{\rm rank(A)}< N$,
all $N \times N$ minors of $A$ vanish identically, 
leading to the trivial solution $\tau_{N,M}(x,y,t) = 0$. 
Therefore, for nontrivial solutions one needs 
$M>N$ and $\mathop{\rm rank}(A)=N$. 
\item
The transformation $A \rightarrow A'=G\,A$ with $G\in \mathrm{GL}(N, \Real)$ 
(corresponding to elementary row operations on~$A$)
amounts to an overall rescaling 
$\tau(x,y,t)\to \tau'(x,y,t)=\det(G)\,\tau(x,y,t)$ 
of the tau-function~\eqref{e:taudet}.
Such rescaling leaves the solution $u(x,y,t)$ in Eq.~\eqref{e:u}
invariant.
This reflects the fact that $N$~independent linear combinations 
of the functions~$f_1,\dots,f_N$ in Eq.~\eqref{e:f}
generate equivalent tau-functions.
This $\mathrm{GL}(N,\Real)$ gauge freedom can be exploited to choose 
the coefficient matrix $A$ in Eq.~\eqref{e:taudet} to be in 
reduced row-echelon form (RREF).
As is well-known, 
the $\mathrm{GL}(N,\Real)$ invariance means that the tau-function~\eqref{e:taudet}
represents a point in the real Grassmannian $\mathrm{Gr}(N,M)$.
\item
Suppose that one of the functions in Eq. \eqref{e:f} contains 
only one exponential term; 
that is, suppose $f_p = a_{p,q}e^{\theta_q}$ with $a_{p,m}=0~\forall m \ne q$. 
Then it is $A(m_1,\dots,m_N) = 0$ whenever $q \notin \{m_1,\dots,m_N\}$, 
and the resulting tau-function~\eqref{e:tauphases} can be expressed as 
$\tau_{N,M}(x,y,t) = e^{\theta_q}\tau'(x,y,t)$,
where $\tau'(x,y,t)$ is a linear combination of exponential
terms containing combinations of $N-1$ distinct phases
chosen from the remaining $M-1$ phases (that is, all $M$ phases but $\theta_q$).
From Eq.~\eqref{e:u} it is evident that $\tau_{N,M}(x,y,t)$ 
and $\tau'(x,y,t)$ generate the same solution of KP.
Moreover, the function $\tau'(x,y,t)$ is 
effectively equivalent to a tau-function $\tau_{N-1,M-1}(x,y,t)$ 
with a coefficient matrix obtained by deleting the $p$-th row 
and $q$-th column of~$A$. 
Hence, the tau-function $\tau_{N,M}(x,y,t)$ is reducible to 
another tau-function $\tau_{N-1,M-1}(x,y,t)$ obtained from a 
Wronskian of $N-1$ functions with $M-1$ distinct phases.
\end{enumerate}

In accordance with the above remarks, throughout this work 
we consider the coefficient matrix~$A$ to be in RREF.
Also, to avoid trivial and singular cases, from now on 
we assume that $M>N$ and $\mathop{\rm rank(A)}=N$,
and that all non-zero $N\times N$ minors of~$A$ are positive.
Finally, we assume that~$A$ 
satisfies the following irreducibility conditions:
\begin{definition}
(Irreducibility)~
A matrix~$A$ of rank $N$ is said to be irreducible if, in RREF:
\begin{enumerate}
\itemsep 0pt
\parsep 0pt
\item
Each column of $A$ contains at least one non-zero element.
\item
Each row of $A$ contains at least one non-zero element in addition
to the pivot.
\end{enumerate}
\label{A:D1}
\end{definition}
\par\kern-2\medskipamount\noindent
Condition~(i) in Definition~\ref{A:D1} requires that each 
exponential phase appear in at least one of the functions $f_1,\dots,f_N$;
condition~(ii) requires that each function contains at least 
two exponential phases.
The reason for condition~(i) should be obvious, for
if~$A$ contains a zero column, the corresponding phase is absent 
from the tau-function, which can then be re-expressed in terms of
an irreducible $N\times(M-1)$ matrix.
The reason for condition~(ii) is to avoid reducible situations 
like those in part~(v) of the above remarks.
Note also that if an $N\times M$ matrix $A$ is irreducible,
then $M>N$.

\subsection{Dominant phase combinations and index pairs}

We now study the asymptotic behavior of the tau-function 
in the $xy$-plane for large values of $|y|$ and finite values of~$t$.  
Let $\Theta$ denote the set of all phase combinations
$\theta_{m_1,\dots,m_N}$ such that $A(m_1,\dots,m_N) \ne 0$,
that is, the set of phase combinations that are actually present
in the tau-function~$\tau(x,y,t)$.

\begin{definition}
(Dominant phase)~~
A given phase combination $\theta_{m_1,\dots,m_N}\in\Theta$ 
is said to be dominant 
for the tau-function $\tau(x,y,t)$ of Eq.~\eqref{e:tauphases}
in a region $R\in\Real^3$ 
if $\theta_{m'_1,\dots,m'_N}\!(x,y,t)\le \theta_{m_1,\dots,m_N}\!(x,y,t)$
for all $\theta_{m'_1,\dots,m'_N}\in\Theta$
and for all $(x,y,t)\in R$.
The region~$R$ is called the dominant region of~$\theta_{m_1,\dots,m_N}$.
\label{D:dominantphase}
\end{definition}

As the phase combinations~$\theta_{m_1,\dots,m_N}(x,y,t)$ 
are linear functions of $x,y$ and~$t$,
each of the inequalities in Definition~\ref{D:dominantphase}
defines a convex subset of~$\Real^3$.
The dominant region~$R$ associated to each phase combination
is also convex,
since it is given by the intersection of finitely many convex subsets.
Furthermore, since the phase combinations are defined globally on~$\Real^3$, 
each point $(x,y,t)\in\Real^3$ belongs to some dominant region~$R$. 
As a result, 
we obtain a partition of the entire~$\Real^3$ into a finite number
of convex dominant regions, 
intersecting only at points on the boundaries of each region.
It is important to note that such boundaries always exist whenever
there is more than one phase combination in the tau-function,
because then there are more than one dominant region in~$\Real^3$.
The significance of the dominant regions 
lies in the following:
\begin{lemma}
The solution $u(x,y,t)$ of the KP equation generated by the
tau-function~\eqref{e:tau} is exponentially small at all points
in the interior of any dominant region.
Thus, the solution is localized only at 
the boundaries of the dominant regions, where a balance exists 
between two or more dominant phase combinations in the 
tau-function of Eq.~\eqref{e:tauphases}. 
\label{L:dominantbalance}
\end{lemma}

\begin{proof}
Let~$R$ be the dominant region associated to~$\theta_{m_1,\dots,m_N}$,
which is therefore the \textit{only} dominant phase in the interior of~$R$.
Then from Eq.~\eqref{e:tauphases} we have that
$\tau_{N,M}(x,y,t)\sim O(e^{\theta_{m_1,\dots,m_N}})$ 
in the interior of $R$. 
As a result, $\log\,\tau_{N,M}(x,y,t)$ locally becomes a 
linear function of $x$ apart from exponentially small terms. 
Then it follows from Eq.~\eqref{e:u} that
the solution $u(x,y,t)$ of KP will be exponentially small 
at all such interior points.~\unskip
\end{proof}

\begin{figure}[t!]
\newdimen\figwidth \figwidth 0.375\textwidth
\centerline{\epsfxsize0.975\figwidth\raise1ex\hbox{\epsfbox{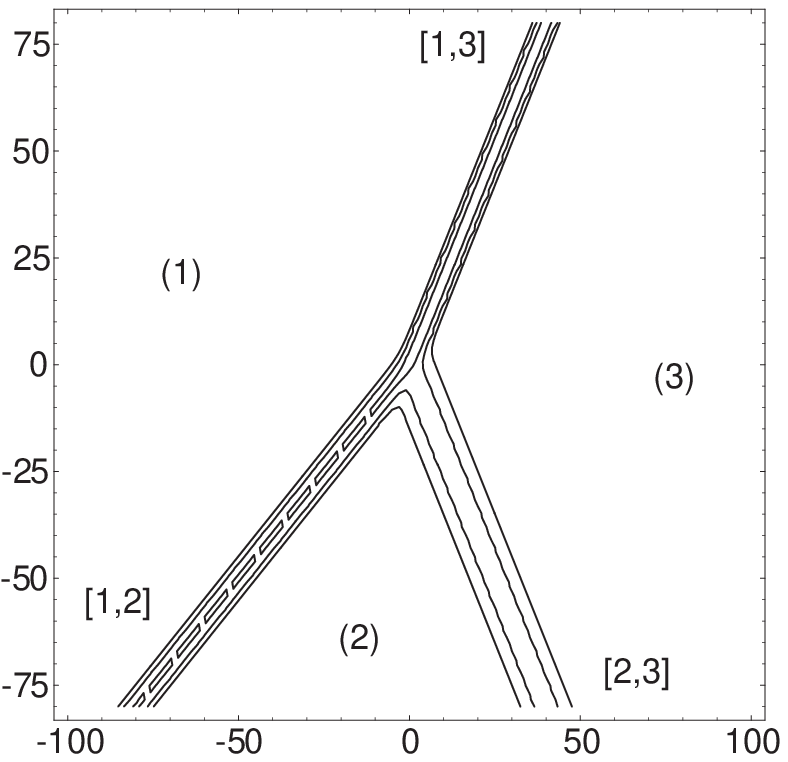}}\quad
\epsfxsize\figwidth\raise1ex\hbox{\epsfbox{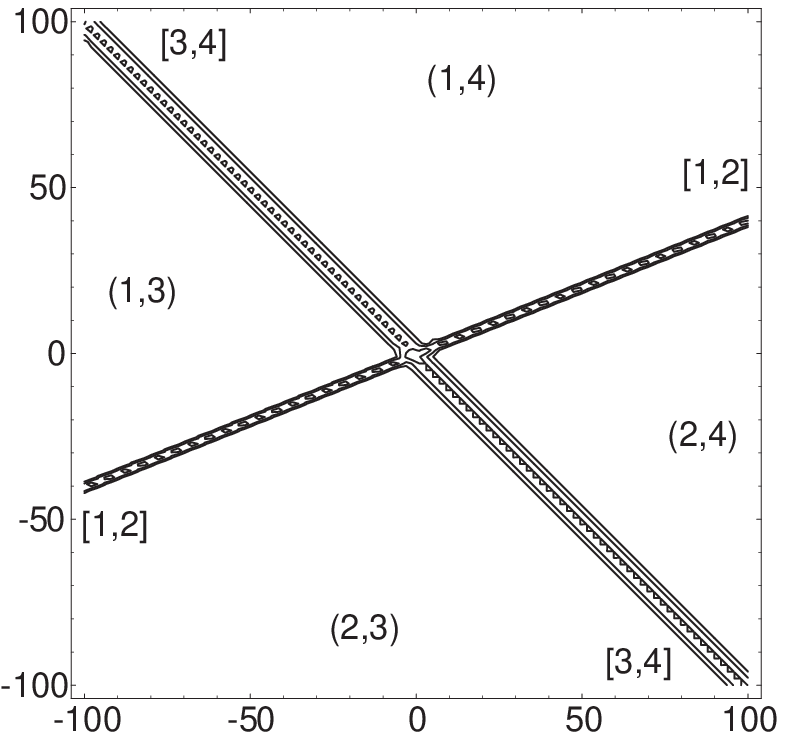}}}
\caption{Dominant phase combinations in the different regions
 of the $xy$-plane (labeled by the indices in parentheses)
 and the asymptotic line solitons (labeled by the indices in square braces)
 for two different line soliton solutions:
 (a)~a fundamental Miles resonance (Y-junction) produced by the tau-function
 with $N=1$, $M=3$ and $(k_1,k_2,k_3)=(-1,0,\frac12)$
 at $t=0$;
 (b)~an ordinary two-soliton solution, produced by the coefficient
 matrix in Example~\ref{x:A2sol} with $(k_1,\dots,k_4)=(-\frac32,-\frac12,0,1)$
 at $t=0$
 (see text for details).
 Here and in all of the following figures, 
 the horizontal and vertical axes are respectively $x$~and~$y$, and 
 the graphs show contour lines of $\log u(x,y,t)$ at a fixed value of~$t$.}
\label{f:dominantphases}
\end{figure}

The boundary between any two adjacent dominant regions is the set of points 
across which \textit{a transition} from one dominant phase combination 
$\theta_{m_1,\dots,m_N}$ to another dominant phase combination 
$\theta_{m'_1,\dots,m'_N}$ takes place. 
Such boundary is therefore identified by the equation
$\theta_{m_1,\dots,m_N} = \theta_{m'_1,\dots,m'_N}$,
which defines a line in the $xy$-plane for fixed values of $t$.
The simplest instance of a transition between dominant phase combinations 
arises for the one-soliton solution~\eqref{e:onesoliton}, 
which is localized along the line $\theta_1=\theta_2$ defining 
the boundary of the two regions of the $xy$-plane where 
$\theta_1$ and $\theta_2$ dominate.
In the one-soliton case, these two regions are simply half-planes,
but in the general case the dominant regions are more complicated, 
although the solution $u(x,y,t)$ is still localized along the boundaries 
of these regions, corresponding to similar phase transitions.
For example, Fig.~\ref{f:dominantphases}a 
illustrates a $(2,1)$-soliton known as 
a \textit{Miles resonance} \cite{JFM1977v79p171} 
(also called a Y-junction),
generated by the tau-function 
$\tau_{1,2}= \e^{\theta_1}+\e^{\theta_2}+\e^{\theta_3}$.
In this case, the $xy$-plane is partitioned into three dominant regions
corresponding to each of the dominant phases $\theta_1$, $\theta_2$ 
and~$\theta_3$. 
Once again,
the solution $u(x,y,t)$ is exponentially small in the interior of each
dominant regions, and is localized along the phase transition boundaries: 
here, 
$\theta_1=\theta_2$,  $\theta_1=\theta_3$ and $\theta_2=\theta_3$.
It should also be noted that some of these regions have infinite extension 
in the $xy$-plane, while others are bounded,
as in the case of resonant soliton solutions,
described in section~\ref{s:examples} and Ref.~\cite{jphysa36p10519}.
Each phase transition which occurs asymptotically as $y\to\pm\infty$
defines an \textit{asymptotic line soliton},
which is infinitely extended in the $xy$-plane.

When studying the asymptotics of the tau-function for large~$|y|$
it is useful to employ coordinate frames 
parametrized by the values of direction~$c$.
That is, we consider the limit $y\to\pm\infty$ along the straight lines
\begin{equation}
L_c:x+cy=\xi\,.
\label{e:Lcdef}
\end{equation}
Note that $c$ increases counterclockwise, namely
from the positive $x$-axis to the negative $x$-axis for $y>0$ 
and from the negative $x$-axis to the positive $x$-axis for $y<0$.
From Eqs.~\eqref{e:theta} and~\eqref{e:Lcdef}, 
the exponential phases along $L_c$ are 
$\theta_m= k_m\,(k_m-c)\,y + k_m\xi + k_m^3t+\theta_{m,0}$.
The difference between two such phases along $L_c$ is then given by
\begin{subequations}
\begin{equation}
\theta_m-\theta_{m'} = 
  (k_m-k_{m'})(k_m+k_{m'}-c)y + (k_m-k_{m'})\xi + (k_m^3-k_{m'}^3)t + \theta_{m,0}-\theta_{m',0}\,,
\label{e:thetaij}
\end{equation}
and the difference between any two phase combinations along $L_c$ is
given by
\begin{equation}
 \theta_{m_1,\dots,m_N} - \theta_{m'_1,\dots,m'_N} =
 \left( \sum_{j=1}^N \big(k_{m_j}-k_{m'_j}\big)\big(k_{m_j}+k_{m'_j}-c\big)
 \right )y
  + \delta(\xi,t)\,,
\label{e:dominantphase}
\end{equation}
where
$\delta(\xi,t)=
  \sum\nolimits_{j=1}^N\big[(k_{m_j}-k_{m'_j})\xi + 
\big(k_{m_j}^3-k_{m'_j}^3\big)t+\theta_{m_j,0}-\theta_{m'_j,0}\big]\,$.
\end{subequations}
In particular, the \textit{single-phase-transition} line 
$L_{m,m'}:\theta_m=\theta_{m'}$,
which will play an important role below, 
is given by Eq.~\eqref{e:Lcdef} with $c_{m,m'}=k_m+k_{m'}$.

Before proceeding further, we introduce the following notations
which will be employed throughout this article.
We denote by $A[m]\in\Real^N$ the $m$-th column of~$A$, 
and we denote by $A[m_1,\dots,m_r]$ the $N\times r$ submatrix 
obtained by selecting the $r$ columns $A[m_1],\dots,A[m_r]$.
We also label the $N$ pivot columns of an irreducible 
$N\times M$ coefficient matrix~$A$ by $A[e_1],\dots,A[e_N]$,
with $1=e_1<e_2<\cdots<e_N< M$,
and we label the $M-N$ non-pivot columns by $A[g_1],\dots,A[g_{M-N}]$,
where $1<g_1<g_2<\cdots<g_{M-N}=M$.
Note that $A$ has $N$ pivot columns because it is rank $N$;
also, $e_1=1$ since $A$ is in RREF, and $e_N<M$ since it is irreducible.
We now establish a result that will be useful in order to 
characterize the asymptotics of the tau-function.
\begin{theorem}
(Single-phase transition)~
Asymptotically as $y\to\pm\infty$,
and for generic values of the phase parameters $k_1,\dots,k_M$,
the dominant phase combinations in the tau-function~\eqref{e:tauphases}
exhibit the following behavior in the $xy$-plane:  
\par\kern-\medskipamount
\begin{enumerate}
\advance\itemsep-4pt
\item 
the set of dominant phase combinations 
remains invariant in time for finite values of $t$.
\item 
the dominant phase combinations in any two adjacent dominant regions 
contain $N-1$ common phases.
\end{enumerate}
\label{T:transition}
\end{theorem}
\vspace{-0.2 in}
The proof of Theorem~\ref{T:transition} is given in the
Appendix.

Consider the single-phase transition as $y\to\pm\infty$
in which a phase $\theta_i$ from the dominant phase combination 
in one region is replaced by another phase $\theta_j$ to produce 
the dominant phase combination in the adjacent region.
We refer to this transition as an $i \to j$ transition,
which takes place along the line $L_{ij}:\, \theta_i = \theta_j$ 
whose direction in the $xy$-plane is given by $c_{ij}=k_i+k_j$.
As $y\to\infty$, it is clear from Eq.~\eqref{e:thetaij} that,
if $k_i<k_j$, 
the transition $i\to j$ takes place from the left of the line $L_{i,j}$ 
to its right, while if $k_i>k_j$ the transition $i\to j$
takes place from the right of the line $L_{i,j}$ to its left.
Thus, as $y\to\infty$,
each dominant phase region $R$ is bounded on the left 
by the transition line $L_{i,j}$ given by to the 
\textit{minimum} value of $c_{i,j}$ that corresponds to an allowed transition, 
and, similarly, on the right by the transition line $L_{i,j}$ given by  
the \textit{maximum} value of $c_{i,j}$ that corresponds to 
an allowed transition. 
Here, an \textit{allowed} transition from one dominant phase combination 
to another means that the minors associated with those phase combinations 
in the tau-function of Eq.~\eqref{e:tauphases}, are both non-zero.
In turn, these non-vanishing minors determine the values of 
$c_{ij}$ corresponding to the allowed single-phase transitions.
A similar statement can be made for transitions occurring as $y\to-\infty$.
So, each dominant phase region~$R$ as $y\to\pm\infty$ has boundaries
defined by a counterclockwise and a clockwise single-phase transitions
which can be identified as follows: 
\begin{corollary}
\label{C:transition}
Suppose that $\theta_{m_1,\dots,m_N}$ is the dominant phase combination
on a region $R$ asymptotically as $y\to\pm\infty$. 
Let~$J$ be the complement of the index set 
$\{m_1,m_2,\dots,m_N\}$ in~$\{1,2,\dots,M\}$.
Also, for each element $j\in J$, 
define a corresponding~$I_j \subseteq\{m_1,m_2,\dots,m_N\}$ 
as the set of all indices $m_r \in \{m_1,m_2,\dots,m_N\}$ 
such that the minor $A(m_1,\dots,m_{r-1},j,m_{r+1},\dots,m_N)\ne0$.
Then:
\begin{enumerate}
\advance\itemsep-4pt
\item
as $y \to \infty$,
the directions of the counterclockwise and clockwise transition
boundaries of~$R$ are respectively given by 
\begin{subequations}
\label{e:boundary}
\begin{equation}
c_+= \min_{i\in I_j, j\in J}[c_{i,j}]\quad\mathrm{with}~k_i>k_j, 
\qquad
c_-= \max_{i \in I_j, j\in J}[c_{i,j}]\quad\mathrm{with}~k_i< k_j\,.  
\end{equation}
\label{e:boundaryA}
\item
as $y \to -\infty$, 
the directions of the counterclockwise and clockwise transition
boundaries of~$R$ are respectively given by 
\begin{equation}
c_+= \min_{i\in I_j, j\in J}[c_{i,j}]\quad\mathrm{with}~k_i<k_j,
\qquad
c_-= \max_{i \in I_j, j\in J}[c_{i,j}]\quad\mathrm{with}~k_i> k_j\,.
\label{e:boundaryB}
\end{equation}
\end{subequations}
\end{enumerate}
\end{corollary}

The results of Theorem~\ref{T:transition} and Corollary~\ref{C:transition}
can now be used to determine the asymptotic behavior of the tau-function
of Eq.~\eqref{e:tau}, 
thereby obtaining an important characterization of the asymptotic line solitons
corresponding to $(N_-,N_+)$-soliton solutions of the KPII equation.
Namely, for the tau-function $\tau_{N,M}(x,y,t)$ of Eq.~\eqref{e:tauphases} 
with generic values of the phase parameters $k_1,\dots,k_M$ 
we have the following:
\par\kern-\medskipamount
\begin{enumerate}
\advance\itemsep-4pt
\item
As $y\to\pm\infty$, the dominant phase combinations of the
tau-function in adjacent regions of the $xy$-plane contain 
$N-1$~common phases and differ by only a single phase.
The transition between any two such dominant phase combinations 
$\theta_{i,m_2,\dots,m_N}$ and $\theta_{j,m_2,\dots,m_N}$
occurs along the line $L_{i,j}:\theta_i=\theta_j$, 
where a single phase $\theta_i$ in the dominant phase combination 
is replaced by a phase $\theta_j$.
Moreover, if the dominant phase combination~$\theta_{i,m_2,\dots,m_N}$ 
in a given region is known, the transition line $L_{i,j}$ 
and the dominant phase combination~$\theta_{j,m_2,\dots,m_N}$
are determined via Corollary~\ref{C:transition}.
In particular, Eqs.~\eqref{e:boundary} for $c_{\pm}$ 
determine explicitly the pair of phase parameters~$k_i$ and~$k_j$ 
corresponding to the single-phase transition $i \to j$ 
across each boundary~$L_{i,j}$ of a given dominant phase region.
\item
As $y\to\pm\infty$ along the line $L_{i,j}$,
the asymptotic behavior of the tau-function is determined by
the balance between the two dominant phase combinations 
$\theta_{i,m_2,\dots,m_N}$ and~$\theta_{j,m_2,\dots,m_N}$,
and is given by
\begin{subequations}
\begin{equation}
\tau_{N,M}\!(x,y,t)\sim 
  V(i,m_2,\dots,m_N)A(i,m_2,\dots,m_N)\,\e^{\theta_{i,m_2,\dots,m_N}}
  + V(j,m_2,\dots,m_N)A(j,m_2,\dots,m_N)\,\e^{\theta_{j,m_2,\dots,m_N}},
\label{e:tauasymp} 
\end{equation}
where 
$V(m_1,\dots,m_N)$ is the Van~der~Monde determinant 
defined in Eq.~\eqref{e:vandermonde},
and where the minors
$A(i,m_2,\dots,m_N)$ and $A(j,m_2,\dots,m_N)$ 
of the coefficient matrix~$A$ are both non-zero.
The solution $u(x,y,t)$ of the KPII equation in a neighborhood of such
a single-phase transition is then obtained from Eq.~\eqref{e:u} as,
\begin{equation}
u(x,y,t) \sim
  \half(k_i-k_j)^2\sech^2\big[\half(\theta_i-\theta_j)\big]\,.
\label{e:asympsol}
\end{equation}
\end{subequations}
Moreover, Lemma~\ref{L:dominantbalance} and Theorem~\ref{T:transition}
together imply that the solution of the KPII equation is exponentially small
everywhere in the $xy$-plane except at the locations of such
single-phase transitions. Equation~\eqref{e:asympsol},
which is a traveling wave solution satisfying the
dispersion relation in Eq.~\eqref{e:dispersionrelation},
coincides with the one-soliton solution in Eq.~\eqref{e:onesoliton}.
Thus, it defines an asymptotic line soliton 
associated with the single-phase transition $i \to j$. 
The phase parameters $k_i$ and $k_j$ associated 
with the single-phase transition $i\to j$
are determined by Eqs.~\eqref{e:boundary};
the soliton amplitude is thus given by $a_{i,j}=|k_i-k_j|$,
and the soliton direction is given by the direction of $L_{i,j}$, 
which is~$c_{i,j}= k_i+k_j$.
\item
All of the asymptotic line solitons resulting from single-phase transitions 
such as the one described above are invariant in time,
in the sense that their number, amplitudes and directions are constants.
\end{enumerate}

Motivated by these results,
\textit{we label each asymptotic line soliton by the index pair~$[i,j]$ 
which uniquely identifies the phase parameters~$k_i$ and $k_j$
in the ordered set $\{k_1,\dots,k_M\}$}.
The results summarized in the above remarks
can be applied to explicitly delineate the dominant phase combinations and 
the asymptotic line solitons associated with the tau-function of a given 
$(N_-,N_+)$-soliton solution of the KPII equation, 
as illustrated by the following example.

\begin{example}
\label{x:A2sol}
When $N=2$ and $M=4$,
Lemma~\ref{L:binetcauchy} implies that the tau-function $\tau(x,y,t)$
is given by
\begin{equation}
\tau(x,y,t)= \Wr(f_1,f_2)= 
  \sum_{1\le m<m'\le 4}(k_{m'}-k_m)\,A(m,m')\,\e^{\theta_m+\theta_{m'}}\,,
\label{e:tau24}
\end{equation}
where the four phases are given by 
$\theta_m= k_mx +k_m^2y+k_m^3t+\theta_{m,0}$ for $m=1,\dots,4$,
as in Eq.~\eqref{e:theta}, 
and where the phase parameters are
ordered as $k_1<\cdots<k_4$. 
We consider the line-soliton solution 
constructed from the two functions $f_1= \e^{\theta_1}+\e^{\theta_2}$
and $f_2=\e^{\theta_3}+\e^{\theta_4}$,
so that the associated $2\times4$ coefficient matrix is
\begin{equation}
A= \begin{pmatrix}1 &1 &0 &0\\ 0 &0 &1 &1\end{pmatrix}.
\label{e:A2sol}
\end{equation}
Then $A(1,2)=A(3,4)=0$, and the remaining four minors are all equal to one.
We apply Corollary~\ref{C:transition} to determine the asymptotic
line solitons associated with the tau-function in Eq.~\eqref{e:tau24}.
First note from the expression
$\theta_{m,m'}=
  (k_m+k_{m'})x+k_m^2+k_{m'}^2)y+(k_m+k_{m'}^3)t+\theta_{m,0}+\theta_{m',0}$
that for every finite value of~$y$
the dominant phase combination as $x\to-\infty$ is given by $\theta_{1,3}$,
which corresponds to the \emph{minimum} value of $k_m+k_{m'}$
such that $A(m,m')\ne0$ (cf.\ Definition~\ref{D:dominantphase}).
Let us denote by $R_{1,3}$ the region of the $xy$-plane where 
$\theta_{1,3}$ is the dominant phase.
The transition boundaries of $R_{1,3}$ are determined by applying 
Corollary~\ref{C:transition} as follows:
The complement of the index set $\{1,3\}$ is $J=\{2,4\}$.
When $j=2\in J$, we have $A(1,2)=0$ but $A(2,3)\ne0$; hence $I_2=\{1\}$.
Similarly, when $j=4$ we have $I_4=\{3\}$ because $A(1,4)\ne0$ but $A(4,3)=0$.
Thus the possible transitions $i\to j$ from $R_{1,3}$ are
$1\to2$ and $3\to4$.
As $y\to\infty$,
the second of Eqs.~\eqref{e:boundaryA} implies that the clockwise
transition boundary of $R_{1,3}$ is given by the transition line~$L_{3,4}$,
whose direction $c_{3,4}=k_3+k_4$ is greater than the direction $c_{1,2}=k_1+k_2$
of the line~$L_{1,2}$.
Across the transition line $L_{3,4}$,
the dominant phase combination switches from $\theta_{1,3}$ to~$\theta_{1,4}$, 
onto the corresponding dominant region, which we denote~$R_{1,4}$.
Similarly, as $y\to-\infty$, 
the first of Eqs.~\eqref{e:boundaryB} implies that the counterclockwise
transition boundary of $R_{1,3}$ is given by the transition line~$L_{1,2}$,
whose direction~$c_{1,2}$ is less than the direction~$c_{3,4}$ of the line $L_{3,4}$.  
This implies that the dominant phase combination and dominant region
change to $\theta_{2,3}$ and $R_{2,3}$, respectively.
Applying Corollary~\ref{C:transition} again to the region~$R_{2,3}$
as $y\to-\infty$,
one finds $J=\{1,4\}$ with $I_1=\{2\}$ and $I_4=\{3\}$,
so the possible transitions from $R_{2,3}$ are $2\to1$ and $3\to4$.
The $2\to1$ transition corresponds to a clockwise transition from
$R_{2,3}$ back to $R_{1,3}$, whereas the $3\to4$ transition corresponds
to a counterclockwise transition from $R_{2,3}$ to the region $R_{2,4}$,
where $\theta_{2,4}$ is the dominant phase combination.
Continuing counterclockwise from $R_{1,3}$ we finally obtain 
the following dominant phase regions asymptotically as $y\to\pm\infty$,
together with the associated single-phase transitions:
\begin{equation}
\def\mapto#1#2{\mathop{\longrightarrow}\limits^{#1\to#2}}
R_{1,3} \mapto12 R_{2,3} \mapto34 R_{3,4} \mapto21 R_{1,4} \mapto43 R_{1,3}\,.
\end{equation}
It is then clear that there are two asymptotic line solitons as $y\to-\infty$ 
as well as $y\to\infty$, and in both cases they 
correspond to the lines $\theta_1=\theta_2$ and $\theta_3=\theta_4$.
The dominant phase regions, denoted by indices $(m,m')$,
and the asymptotic line solitons, identified by the index pairs $[i,j]$,
are illustrated in Fig.~\ref{f:dominantphases}b.
\end{example}

In the following section we obtain several results that will allow us 
to identify more precisely the index pairs corresponding to each 
asymptotic line soliton. 
In addition, we will prove a general result concerning the \textit{numbers}
of asymptotic line solitons present in any 
$(N_-, N_+)$-soliton solution corresponding to a tau-function 
with an arbitrary number of functions~$f_1,\dots,f_N$
and arbitrary linear combinations of the exponential phases 
$e^{\theta_1},\dots,e^{\theta_M}$ in each function.

\begingroup
\def\sum{\mathop{\textstyle\truesum}}
\def\span{\mathop{\rm span}\nolimits}
\let\~=\tilde

\section{Asymptotic line solitons and the coefficient matrix}
\label{s:asymptotics}

In this section we continue our investigation of the tau-function in
the general setting introduced in section~\ref{s:general}.
We have seen in the previous section that 
an asymptotic line soliton corresponds to a dominant balance 
between two phase combinations in the tau-function. 
But we still need to identify which phase combinations 
in a given tau-function are indeed dominant as $y\to\pm\infty$.
This requires a detailed study of the structure of the 
$N \times M$ coefficient matrix $A$~associated with the tau-function. 
In this section we carry out this analysis, which enables us 
to explicitly identify all the asymptotic line solitons
of a given tau-function in an algorithmic fashion.
One of our main results of this section will be to establish that, 
for arbitrary values of~$N$ and~$M$,
and for irreducible coefficient matrices (cf.\ Definition~\ref{A:D1}) 
with non-negative $N \times N$ minors,
the tau function~\eqref{e:tau} produces an $(N_-,N_+)$-soliton solution
with $N_-=M-N$ and $N_+=N$,
i.e., a solution in which there are 
$N_-=M-N$ asymptotic line solitons as $y\to-\infty$
and $N_+=N$ asymptotic line solitons as $y\to\infty$.

\subsection{Dominant phases and structure of the coefficient matrix}
\label{s:rankconditions}

We begin by presenting a simple yet useful result that will be
frequently used to determine the dominant phase combinations in the
tau-function as $y\to\pm\infty$.
\begin{lemma}
(Dominant phase conditions)~ 
As $y\to\pm\infty$ along the line $L_{i,j}:\theta_i=\theta_j$
with $i<j$,
the exponential phases $\theta_1,\dots,\theta_M$ satisfy 
the following relations. 
\begin{enumerate}
\par\kern-0.5\medskipamount
\itemsep 0pt
\parsep 0pt
\item 
As $y\to\infty$,\,
$\theta_m<\theta_* ,\, \forall \, m\in\{i+1,\dots,j-1\}$,
and $\theta_m>\theta_* ,\, \forall \, m\in\{1,\dots,i-1,j+1,\dots,M\}$,
where $\theta_*:=\theta_i=\theta_j$.
\item
As $y\to-\infty$,\,
$\theta_m>\theta_*,\, \forall \, m\in\{i+1,\dots,j-1\}$, 
while $\theta_m<\theta_*,\, \forall \, m\in\{1,\dots,i-1,j+1,\dots,M\}$.
\end{enumerate}
\par\kern-\smallskipamount\noindent
\label{A:dominantphase}
\end{lemma}
\begin{proof}
It follows from Eq.~\eqref{e:thetaij} that, along the line $L_{i,j}$,
the difference beetween any two exponential phases $\theta_m$ and
$\theta_{m'}$ is given by 
\begin{equation}
\theta_m - \theta_{m'}= 
  (k_m-k_{m'})[(k_m+k_{m'})-(k_i+k_j)]y + \delta'(\xi,t)\,,
\label{e:thetamn}
\end{equation}
where $\delta'(\xi,t)$ is a linear function of $\xi$ and~$t$ and 
which also depends on the constants $\theta_{m,0}$, $\theta_{m',0}$, 
$\theta_{i,0}$ and~$\theta_{j,0}$,
and where we used the fact that the direction of the line 
$L_{i,j}$ is $c_{i,j}=k_i+k_j$.
It is clear that the sign of $\theta_m-\theta_{m'}$
as $y\to\pm\infty$ and for finite values of $\xi$ and $t$ is determined by 
the coefficient of $y$ in the right-hand side of Eq.~\eqref{e:thetamn} 
Then, setting $m'=i$ (or $m'=j$) in Eq.~\eqref{e:thetamn}
one obtains the desired inequalities.~%
\end{proof}

\begin{figure}[t!]
\medskip
\centerline{\epsfxsize0.675\textwidth\epsfbox{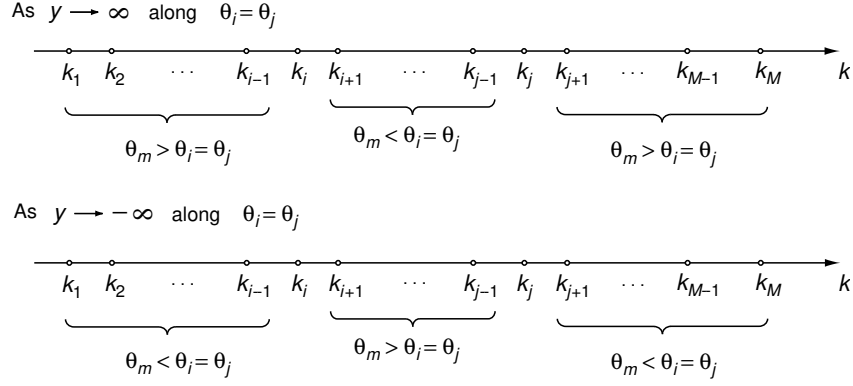}}
\caption{Relations among the exponential phases as $y\to\pm\infty$
along the direction $L_{i,j}:\theta_i=\theta_j$.}
\label{A:fdominant}
\end{figure}

Lemma~\ref{A:dominantphase}, which is illustrated in Fig.~\ref{A:fdominant},
will be used to obtain a set of conditions that are necessary 
for a given pair of phase combinations in the tau-function to be dominant. 
These conditions are given in terms of the vanishing of certain 
$N \times N$ minors of the coefficient matrix~$A$, 
and they determine which phase combinations are present (or absent) 
in the tau-function of Eq.~\eqref{e:tau}.
In order to derive these conditions, it is convenient to introduce 
two submatrices $P_{i,j}$ and $Q_{i,j}$ associated with any index
pair $[i,j]$ with $1\le i<j\le M$, and given by
\begin{equation}
P_{i,j}= A[1,2,\dots,i-1,j+1,\dots,M]\,,\qquad
Q_{i,j}= A[i+1,\dots,j-1]\,.
\label{e:PQ}
\end{equation}
The matrix $P_{i,j}$ contains the consecutive columns of~$A$ 
to the left of column~$A[i]$ and those to the right of column~$A[j]$, 
while $Q_{i,j}$ contains the consecutive columns of~$A$ 
between columns $A[i]$ and~$A[j]$.
Using the matrices $P_{i,j}$ and $Q_{i,j}$ and the dominant phase
conditions in Lemma~\ref{A:dominantphase} we then have:
\begin{lemma}
(Vanishing minor conditions)~
Suppose that the index pair $[i,j]$ identifies an asymptotic 
line soliton. Let the two dominant phase combinations 
along the line $L_{i,j}:\theta_{i}=\theta_{j}$
be given by\,\break
$\theta_{i,p_1,\dots,p_r,q_1,\dots,q_s}$\, and\, 
$\theta_{j,p_1,\dots,p_r,q_1,\dots,q_s}$, \, and let 
$A(i,p_1,\dots,p_r,q_1,\dots,q_s)$\,, 
$A(j,p_1,\dots,p_r,q_1,\dots,q_s)$\, be the corresponding 
non-zero minors
where $A[p_1],\dots,A[p_r]\in P_{i,j}$\, and\, 
$A[q_1],\dots,A[q_s]\in Q_{i,j}$.
\begin{enumerate}
\item
If $[i,j]$ identifies an asymptotic line soliton as $y\to\infty$, then 
\par\kern-\medskipamount
\begin{enumerate}
\itemsep 0pt
\parsep 0pt
\item
all $N \times N$ minors obtained by replacing one of
the columns $A[i],A[j],A[q_1],\ldots,A[q_s]$ from either 
$A(i,p_1,\dots,p_r,q_1,\dots,q_s)$ or~$A(j,p_1,\dots,p_r,q_1,\dots,q_s)$
with any column $A[p] \in P_{i,j}$, are zero;
\item
all $N \times N$ minors obtained by replacing one of the 
columns $A[q_1],\dots,A[q_s]$ from either \break
$A(i,p_1,\dots,p_r,q_1,\dots,q_s)$ or $A(j,p_1,\dots,p_r,q_1,\dots,q_s)$
with either $A[i]$ or $A[j]$, are zero.
\end{enumerate}
\item
If $[i,j]$ identifies an asymptotic line soliton as $y\to-\infty$, then
\par\kern-\medskipamount
\begin{enumerate}
\itemsep 0pt
\parsep 0pt
\item
all $N \times N$ minors obtained by replacing one of the
columns $A[i],A[j],A[p_1],\ldots,A[p_r]$ from either
$A(i,p_1,\dots,p_r,q_1,\dots,q_s)$ or $A(j,p_1,\dots,p_r,q_1,\dots,q_s)$
with any column $A[q] \in Q_{i,j}$, are zero;
\item
all $N \times N$ minors obtained by replacing one of the
columns $A[p_1],\dots,A[p_r]$ from either \break
$A(i,p_1,\dots,p_r,q_1,\dots,q_s)$ or $A(j,p_1,\dots,p_r,q_1,\dots,q_s)$
with either $A[i]$ or $A[j]$, are zero.
\end{enumerate}
\end{enumerate}
\label{L:zerominors}
\end{lemma}
\begin{proof}
All of the above conditions follow from the repeated use of the dominant
phase conditions in Lemma~\ref{A:dominantphase}. 
For example, as $y \to \infty$ along the line $L_{i,j}$,
Lemma~\ref{A:dominantphase} implies 
$\theta_p > \theta_m$ for all $p \in \{1,\ldots,i-1,j+1,\ldots,M\}$
and for all $m \in \{i,j,q_1,\ldots,q_s\}$.
Consequently, if condition~(b) in part~(i) of the Lemma does not hold, 
each of the phase combinations obtained by replacing
$\theta_m$ with $\theta_p$
in either $\theta_{i,p_1,\dots,p_r,q_1,\dots,q_s}$
or $\theta_{j,p_1,\dots,p_r,q_1,\dots,q_s}$ 
will be greater than both $\theta_{i,p_1,\dots,p_r,q_1,\dots,q_s}$ and
$\theta_{j,p_1,\dots,p_r,q_1,\dots,q_s}$.
But this contradicts the hypothesis that 
$\theta_{i,p_1,\dots,p_r,q_1,\dots,q_s}$ and
$\theta_{j,p_1,\dots,p_r,q_1,\dots,q_s}$ are the dominant phase combinations
as $y\to\infty$ along $L_{i,j}$.
The other conditions follow in a similar fashion.
\end{proof}

We should emphasize that $[i,j]$ denotes  
an asymptotic line soliton either as $y\to\infty$ or as $y\to-\infty$.
In general, the asymptotic solitons (and therefore the index pairs) 
as $y\to\infty$ and those as $y\to-\infty$ are different. 
Thus, in principle there is no relation among  
the matrices $P_{i,j}$ and $Q_{i,j}$
relative to solitons as $y\to\infty$ and those 
associated to solitons as $y\to-\infty$.

Lemma~\ref{L:zerominors} allows us to determine the ranks of the submatrices 
$P_{ij}$ and $Q_{ij}$ associated with each asymptotic line soliton~$[i,j]$. 
This information will be exploited later in Theorem~\ref{A:T1} 
to identify explicitly the asymptotic line solitons produced
by any given tau-function. 
The next two results are direct consequences of the conditions 
specified in Lemma~\ref{L:zerominors}.
\begin{lemma}
(Span)~
Let $A[p_1],\dots,A[p_r]$ be the columns from $P_{i,j}$
and $A[q_1],\dots,A[q_s]$ be the columns from $Q_{i,j}$
in the minors associated with the dominant pair of phase combinations,
as in Lemma~\ref{L:zerominors}
\begin{enumerate}
\item
If $[i,j]$ identifies an asymptotic line soliton as $y\to\infty$, 
the columns $A[p_1],\dots,A[p_r]$ form a basis for the column space
of the matrix $P_{i,j}$.
\item
If $[i,j]$ identifies an asymptotic line soliton as $y\to-\infty$, 
the columns $A[q_1],\dots,A[q_s]$ form a basis for the column space
of the matrix $Q_{i,j}$.
\end{enumerate}
\label{L:span}
\end{lemma}
\begin{proof}
We prove part~(i).
Since $A(i,p_1,\dots,p_r,q_1,\dots,q_s)\ne0$ by Lemma~\ref{L:zerominors},
the set of columns $\A= \{A[i],A[p_1],\dots,A[p_r],A[q_1],\dots,A[q_s]\}$
is a basis of~$\Real^N$.
Hence the set $\{A[p_1],\dots,A[p_r]\}\subset\A$ is linearly independent.
Moreover, for any $A[p]\in P_{i,j}$ we can expand $A[p]$ with respect to~$\A$:
\begin{equation}
A[p]= 
  a\, A[i] + \sum_{m=1}^r b_m A[p_m] + \sum_{m=1}^s c_m A[q_m]\,.
\label{e:span}
\end{equation}
Replacing one of the columns $A[i],A[q_1],\ldots,A[q_s]$ 
in $A(i,p_1,\dots,p_r,q_1,\dots,q_s)$ with $A[p] \in P_{ij}$,
we have from Lemma~\ref{L:zerominors}.i.a that 
\[
A(p,p_1,\dots,p_r,q_1,\dots,q_s)=0, \qquad
A(i,p_1,\dots,p_r,q_1,\dots,q_{m-1},p,q_{m+1},\dots,q_s)=0\,.
\]
Hence in Eq.~\eqref{e:span} we have $a=0$
and $c_m=0$ $\forall m=1,\dots,s$.
Therefore $A(p)\in\span(A[p_1,\dots,p_r])$
for all $A[p]\in P_{i,j}$.
Similarly, part~(ii) follows from the conditions 
in Lemma~\ref{L:zerominors}.ii.a.
\end{proof}
%
%
\begin{lemma}
(Rank conditions)~
Let $r$ be the number of columns from~$P_{i,j}$ 
and let~$s$ be the the number of columns from~$Q_{i,j}$
in the minors associated with the dominant pair of phase combinations,
as in Lemma~\ref{L:zerominors}.
\begin{enumerate}
\item
If $[i,j]$ identifies an asymptotic line soliton as $y\to\infty$, then\,
$\rank(P_{i,j})=r\le N-1$\, and\,
$\rank(P_{i,j}|A[i])= \rank(P_{i,j}|A[j])= \rank(P_{i,j}|A[i,j])= r+1$.
\item
If $[i,j]$ identifies an asymptotic line soliton as $y\to-\infty$, then\,
$\rank(Q_{i,j})=s\le N-1$\, and\,
$\rank(Q_{i,j}|A[i])\!\break= \rank(Q_{i,j}|A[j])= \rank(Q_{i,j}|A[i,j])= s+1$.
\end{enumerate}
\par\kern-\medskipamount\noindent
Above and hereafter, $(A|B)$ denotes the matrix~$A$ augmented by the matrix~$B$.
\label{L:rank}
\end{lemma}
\begin{proof}
Let us prove part~(i).
Since the columns $A[p_1],\dots,A[p_r]$ form a basis for 
the column space of $P_{i,j}$, 
from Lemma~\ref{L:span}.i we immediately have $\rank(P_{i,j})=r$.
Moreover, since $\A=\{A[i],A[p_1],\dots,A[p_r],\break A[q_1],\dots,A[q_s]\}$
is a basis for $\Real^N$,
the vectors $A[i],A[p_1],\dots,A[p_r]$ are linearly independent,
and therefore $\rank(P_{i,j}|A[i])= r+1$.
Similarly, replacing $A[i]$ with $A[j]$ in the previous statement
we have $\rank(P_{i,j}|A[j])= r+1$.
It remains to prove that 
$\rank(P_{i,j}|A[i,j])=r+1$.
Expanding the $j$-th column of~$A$ in terms of~$\A$ 
as in Lemma~\ref{L:span} we have
\begin{equation}
A[j]= a\, A[i] + \sum_{m=1}^r b_m A[p_m]+ \sum_{m=1}^s c_m A[q_m]\,.
\label{e:jnexp}
\end{equation}
By replacing one of the columns $A[q_1], \ldots A[q_s]$ 
in $A(i,p_1,\dots,p_r,q_1,\dots,q_s)$ with $A[j]$, from
Lemma~\ref{L:zerominors}.i.b we have that 
$A(i,p_1,\dots,p_r,q_1,\dots,q_{m-1},j,q_{m+1},\dots,q_s)=0\,$.
Therefore $c_m=0$ for all $m=1,\dots,s$.
Consequently we have 
$A[j]\in\span(A[i],A[p_1],\dots,A[p_r])$,
which implies that $\rank(P_{i,j}|A[i,j])= r+1$.
Similarly, using Lemma~\ref{L:zerominors}.ii.b
one can establish the corresponding results
in part~(ii) for the asymptotic line solitons as $y\to-\infty$.
\end{proof}

It is important to note that, 
even though Lemmas~\ref{L:span}--\ref{L:rank} were proved
by using the vanishing minor conditions in Lemma~\ref{L:zerominors},
they provide additional information on the structure of 
the coefficient matrix~$A$.
For example, when $r<N-1$ for an asymptotic line soliton as $y\to\infty$, 
Lemma~\ref{L:rank} yields $\rank(P_{i,j}|A[i,j])<N$,
and when $s<N-1$ for an asymptotic line soliton as $y\to-\infty$,
Lemma~\ref{L:rank} yields $\rank(Q_{i,j}|A[i,j])<N$.
As a consequence, we immediately have
the following additional vanishing minor conditions:
\begin{enumerate}
\item
If $[i,j]$ identifies an asymptotic line soliton as $y\to\infty$, then
\begin{subequations}
\label{e:vanishingminors}
\begin{equation}
A(i,j,p_1,\dots,p_r,m_1,\dots,m_{N-r-2})=0\, \qquad
\forall \, \{m_1,\dots,m_{N-r-2}\}\subset\{1,\dots,M\}.
\end{equation}
\item
If $[i,j]$ identifies
an asymptotic line soliton as $y\to-\infty$, then
\begin{equation}
A(i,j,q_1,\dots,q_s,m_1,\dots,m_{N-s-2})=0\,\qquad
\forall \{m_1,\dots,m_{N-s-2}\}\subset\{1,\dots,M\}\,.
\end{equation}
\end{subequations}
\end{enumerate}
It should also be noted that,
when $[i,j]$ identifies an asymptotic line soliton as $y\to\infty$,
Lemma~\ref{L:rank}.i only provides information on $P_{i,j}$, and
the only condition on $Q_{i,j}$ is that $\rank(Q_{i,j})\ge s$.
Similarly, when $[i,j]$ identifies an asymptotic line soliton as $y\to-\infty$,
all we know about $P_{i,j}$ is that $\rank(P_{i,j})\ge r$.

\subsection{Characterization of the asymptotic line solitons from the coefficient matrix}

In section~\ref{s:rankconditions} we derived several conditions
that an index pair $[i,j]$ must satisfy in order to identify an
asymptotic line soliton.
In this section we apply the results developed in section~\ref{s:rankconditions}
to obtain a complete characterization of the incoming and outgoing
asymptotic line solitons of a generic line-soliton solution of the KPII
equation.

\begin{lemma}
(Pivots and non-pivots)~
Consider an index pair $[i,j]$ with $1\le i<j\le M$.
\begin{enumerate}
\item
If $[i,j]$ identifies an asymptotic line soliton as $y\to\infty$, 
the index~$i$ labels a pivot column of the coefficient matrix~$A$.
That is, $A[i]=A[e_n]$ with $1\le n\le N$.
\item
If $[i,j]$ identifies an asymptotic line soliton as $y\to-\infty$, 
the index~$j$ labels a non-pivot column of the coefficient matrix~$A$.
That is, $A[j]=A[g_n]$ with $1\le n\le M-N$.
\end{enumerate}
\label{L:pivot}
\end{lemma}
\begin{proof}
We first prove part~(i).
Suppose that $\theta_{i,m_2,\dots,m_N}$ 
is one of the dominant phase combinations corresponding to the
asymptotic line soliton $[i,j]$ as $y\to\infty$.
The corresponding minor $A(i,m_2,\dots,m_N)$ is non-zero.
Since~$A$ is in RREF,
we have $A[i]= \sum_{r=1}^n c_r A[e_r]$ for some $n\le N$,
where $e_1<\cdots<e_n\le i$.
Therefore $A(i,m_2,\dots,m_N)= \sum_{r=1}^n c_r A(e_r,m_2,\dots,m_N)$.
If $e_n<i$, we have $A[e_1],\dots,A[e_n]\in P_{i,j}$,
where $P_{i,j}$ is the submatrix of~$A$ defined in Eq.~\eqref{e:PQ}.
Then from condition~(a) in Lemma~\ref{L:zerominors}.i we have 
$A(e_r,m_2,\dots,m_N)=0$ $\forall r=1,\dots,n$,
implying that $A(i,m_2,\dots,m_N)=0$.
But this is impossible, since $\theta_{i,m_2,\dots,m_N}$
is a dominant phase combination.
Therefore we must have $i=e_n$, meaning that $A[i]$ is a pivot column.

Part~(ii) follows from the rank conditions in Lemma~\ref{L:rank}.ii.
In particular,  $\rank(Q_{i,j}|A[i])=\break\rank(Q_{i,j}|A[i,j])=s+1$ 
implies that $A[j]\in\span(A[i],\dots,A[j-1])$.
Since $A$ is in RREF, none of its pivot column can be spanned by the preceding columns.
Hence $A[j]$ cannot be a pivot column.
\end{proof}

Lemma~\ref{L:pivot} identifies outgoing and incoming asymptotic line solitons 
respectively with the pivot and the non-pivot columns of~$A$.
It is then natural to ask if in fact each of the $N$ pivot columns 
and each of the $M-N$ non-pivot columns identifies an outgoing or incoming 
line soliton, and whether such identification is unique. 
Both of these questions can be answered affirmatively by the following
theorem which constitutes one of the main results of this work, and
is proved in the Appendix.
\begin{theorem}
(Asymptotic line solitons)~
Let $\tau_{N,M}(x,y,t)$ be the tau-function in Eq.~\eqref{e:taudet} 
associated with a rank $N$, irreducible coefficient matrix~$A$ 
with non-negative minors.
\begin{enumerate}
\item
For each pivot index $e_n$ 
there exists a unique asymptotic line soliton as $y\to\infty$,
identified by an index pair $[e_n,j_n]$ 
with $n=1,\dots,N$ and $1\le e_n<j_n\le M$.
\item
For each non-pivot index $g_n$ 
there exists a unique asymptotic line soliton as $y\to-\infty$,
identified by an index pair $[i_n,g_n]$ 
with $n=1,\dots,M-N$ and $1\le i_n<g_n\le M$.
\end{enumerate}
Thus, the solution of KPII generated by the coefficient matrix~$A$ 
via Eq.~\eqref{e:taudet}
has exactly $N_+=N$ asymptotic line solitons as $y\to\infty$
and $N_-=M-N$ asymptotic line solitons as $y\to-\infty$.
\label{A:T1}
\end{theorem}
Part~(i) of Theorem~\ref{A:T1} uniquely identifies the asymptotic
line solitons as $y\to\infty$ by the index pairs $[e_n,j_n]$ where
$e_n<j_n$. The indices $e_1,\dots,e_N$ label the $N$~pivot columns 
of~$A$, however, the $j_n$'s may correspond to either pivot or non-pivot 
columns, and indeed both cases appear in examples.
Moreover, when the pivot indices are sorted in increasing order
 $1=e_1<e_2<\cdots<e_N<M$, the indices $j_1,\dots,j_N$ in general 
\textit{are not sorted} in any specific order.
For example, the line solitons as $y\to\infty$ generated by the matrix~$A$ 
in Eq.~\eqref{e:A3partialresonance} of section~\ref{s:examples}
have $j_1<j_3<j_2$.\,\  
In fact, the indices $j_1,\dots,j_N$ need not necessarily even be distinct. 
Similarly, part~(ii) of Theorem~\ref{A:T1} uniquely identifies 
the asymptotic line solitons as $y\to-\infty$ by index pairs $[i_n,g_n]$,
where $i_n<g_n$. In this case, the indices $g_1,\dots,g_{M-N}$ label the 
$M-N$ non-pivot columns of~$A$, but the $i_n$'s may
correspond to either pivot or non-pivot columns.
Moreover, when the non-pivot indices are sorted in increasing order 
$1<g_1<\cdots<g_{M-N}=M$), 
the indices $i_1,\dots,i_{M-N}$ are not in general sorted,
and need not be distinct.
Theorem~\ref{A:T1} yields an important characterization of the 
solution via the associated coefficient matrix~$A$.It provides a 
concrete method to identify the 
asymptotic line solitons as $y\to\pm\infty$,
as illustrated with the two examples below.
Further examples are discussed in section~\ref{s:examples}.

\begin{figure}[t!]
\newdimen\figwidth \figwidth 0.375\textwidth
\centerline{\epsfxsize0.9675\figwidth\raise1.5ex\hbox{\epsfbox{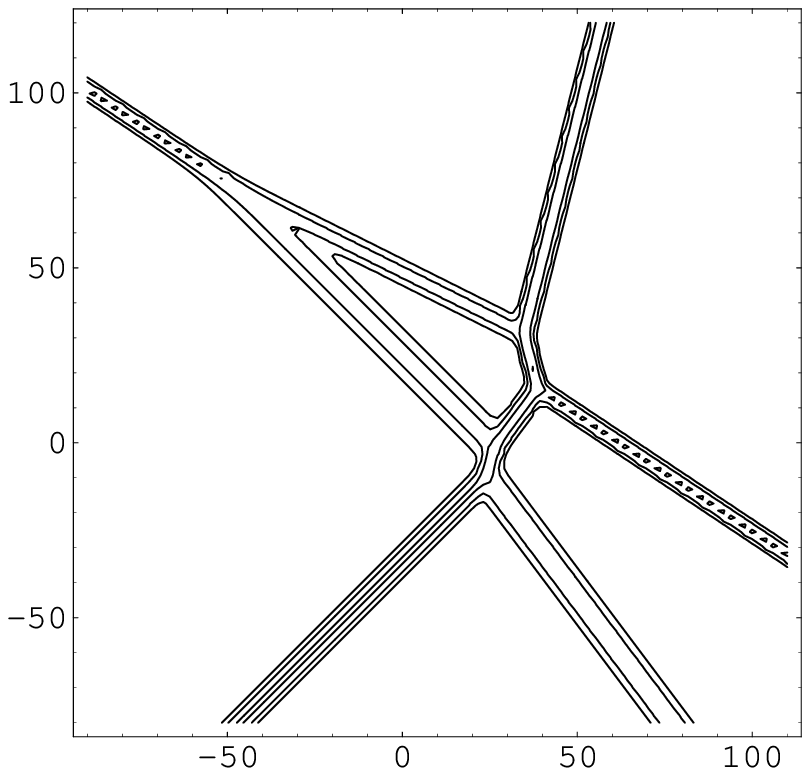}}\quad
\epsfxsize\figwidth\raise1ex\hbox{\epsfbox{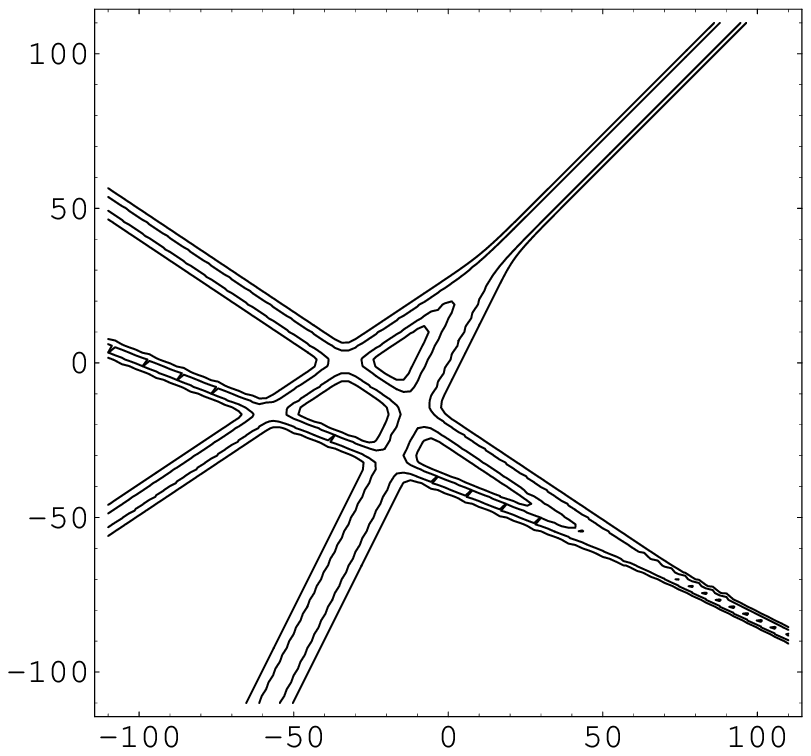}}}
\caption{Line soliton solutions of KPII:
(a)~the (3,2)-soliton solution generated by the coefficient matrix~$A$ 
in Example~\ref{x:3to2} with 
$(k_1,\dots,k_5)=(-1, 0, \frac14, \frac34, \frac54)$ at $t=-32$;
(b)~the inelastic 3-soliton solution 
generated by the coefficient matrix~$A$ in
Example~\ref{x:inelastic3s} 
with $(k_1,\dots,k_6)=(-1, -\frac12, 0, \frac12, 1,\frac32)$ at $t=20$
(see text for details).}
\label{f:x3s}
\end{figure}

\begin{example}
Consider the tau-function $\tau_{N,M}$ with $N=2$ and $M=5$ generated by
the coefficient matrix
\begin{equation}
A= \begin{pmatrix}
1 &1 &0 &\!-1 &\!-2\\ 0 &0 &1 &1 &1
\end{pmatrix}
\end{equation}
The pivot columns of~$A$ are labeled by the indices $\{e_1,e_2\}=\{1,3\}$,
and the non-pivot columns by the indices $\{g_1,g_2,g_3\}=\{2,4,5\}$.
Thus, from Theorem~\ref{A:T1} we know that there will be $N_+=N=2$ asymptotic
line solitons as $y\to\infty$, identified by the index pairs $[1,j_1]$
and $[3,j_2]$ for some $j_1>1$ and $j_2>3$,
and that there will be
$N_-=M-N=3$ asymptotic line solitons as $y\to-\infty$, identified by the
index pairs $[i_1,2]$, $[i_2,4]$ and $[i_3,5]$, for some
$i_1<2$, $i_2<4$ and $i_3<5$.
We first determine the asymptotic line solitons as $y\to\infty$
using part~(i) of Theorem~\ref{A:T1} together with the 
rank conditions~in Lemma~\ref{L:rank}.i.
Then we find the asymptotic line solitons as $y\to-\infty$
using part~(ii) of Theorem~\ref{A:T1} and the rank conditions
in Lemma~\ref{L:rank}.ii.

For the first pivot column, $e_1=1$, we start with $j=2$ and consider
the submatrix
$P_{1,2}= \bigl(\begin{smallmatrix} 0 &\!-1 &\!-2\\ 1 &1 &1
\end{smallmatrix}\bigr)\,$.
Since $\rank(P_{1,2})=2>1=N-1$,
from Lemma~\ref{L:rank}.i we conclude that the pair $[1,2]$ cannot
identify an asymptotic line soliton as $y\to\infty$.
Incrementing $j$ to $j=3,4,5$ and checking the rank of each submatrix
$P_{1,j}$ we find that the rank conditions in Lemma~\ref{L:rank}.i
are satisfied when $j=4$:
$P_{1,4}= \bigl(\begin{smallmatrix}\!-2\\1\end{smallmatrix}\bigr)=A[5]$, 
so $\rank(P_{1,4})=1$ and
$\rank(P_{1,4}|A[1])= \rank(P_{1,4})|A[4])= 2$
(The condition $\rank(P_{1,4}|A[1,4])= 2$ is trivial here,
since any three columns are linearly dependent.)
Thus, the first asymptotic line soliton as $y\to\infty$ is identified
by the index pair $[1,4]$.
For the second pivot, $e_2=3$, proceeding in a similar manner we find that
$j=4$ does not satisfy the rank conditions, since $P_{3,4}$ has rank~2.
But $j=5$ satisfies Lemma~\ref{L:rank}.i, since
$P_{3,5}= \bigl(\begin{smallmatrix} 0 &\!-1 &\!-2\\ 1 &1 &1
\end{smallmatrix}\bigr)\,$,
which yields $\rank(P_{3,5})=1$ and
$\rank(P_{3,5}|A[3])= \rank(P_{3,5})|A[5])= 2$.
(Again, $\rank(P_{3,5}|A[3,5])= 2$ is trivially satisfied here.)
So the asymptotic line solitons as $y\to\infty$ are given by
the index pairs $[1,4]$ and $[3,5]$,
and the associated phase transition diagram
(cf.\ Corollary~\ref{C:transition}) is given by
\begin{equation*}
R_{1,3} \mapto35 R_{1,5} \mapto14 R_{4,5}\,.
\end{equation*}

We now consider the asymptotics for $y\to-\infty$.
Starting with the non-pivot column $g_1=2$, 
the only column to its left is $i=1$.  
We have $Q_{1,2}=\emptyset$, and
$\rank(Q_{1,2}|A[1])= \rank(Q_{1,2}|A[2])= 
\rank(Q_{1,2}|A[1,2])=1$.
Consequently, the pair $[1,2]$ identifies an asymptotic line soliton
as $y\to-\infty$.
For $g_2=4$ we consider $i=1,2,3$ and find that 
the rank conditions in Lemma~\ref{L:rank}.ii are satisfied only for $i=2$: 
in this case,
$Q_{2,4}= \bigl(\begin{smallmatrix}0\\1\end{smallmatrix}\bigr)=A[3]$,
so $\rank(Q_{2,4})= 1=N-1$ and
$\rank(Q_{2,4}|A[2])= \rank(Q_{2,4}|A[4])= 2$,
while $\rank(Q_{2,4}|A[2,4])= 2$ is trivially satisfied.
Hence $[2,4]$ is the unique asymptotic line soliton as $y\to-\infty$
associated to the non-pivot column $g_2=4$.
In a similar way we can uniquely identify the last asymptotic line soliton 
as $y\to-\infty$ as given by the indices $[3,5]$.
The phase transition diagram for $y\to-\infty$ is thus given by
\begin{equation*}
R_{1,3} \mapto12 R_{2,3} \mapto 24 R_{3,4} \mapto35 R_{4,5}\,.
\end{equation*}
To summarize, there are $N_+=2$ outgoing line solitons, 
each associated with one of the pivot columns $e_1=1$ and $e_2=3$,
given by the index pairs $[1,4]$ and $[3,5]$, 
and there are $N_-=3$ incoming line solitons, 
each associated with one of the non-pivot columns $g_1=2$, $g_2=4$ and $g_3=5$,
given by the index pairs $[1,2]$, $[2,4]$ and $[3,5]$.
A snapshot of the solution at $t=-32$ is shown in Fig.~\ref{f:x3s}a.
\label{x:3to2}
\end{example}

\begin{example}
Consider the tau-function with $N=3$ and $M=6$ generated by the coefficient
matrix in RREF
\begin{equation}
A= \begin{pmatrix}
1 &1 &1 &0 &0 &0\\
0 &0 &0 &1 &0 &\!-1\\ 
0 &0 &0 &0 &1 &2
\end{pmatrix}\,.
\label{e:A3inelasticO}
\end{equation}
Again, we first determine the asymptotic line solitons as $y\to\infty$;
then we find the asymptotic line solitons as $y\to-\infty$.

The pivot columns of~$A$ are labeled by the indices $e_1=1$,
$e_2=4$ and $e_3=5$.
Thus, we know that the asymptotic line solitons as $y\to\infty$ 
will be given by the index pairs $[1,j_1]$, $[4,j_2]$ and $[5,j_3]$ for
some $j_1,\dots,j_3$.
Starting with the first pivot, $e_1=1$, we take $j=2,3,\dots$
and check the rank of the submatrix $P_{i,j}$ in each case.
When $j=2$ we have
$P_{1,2}= \left(\!
\begin{smallmatrix} 
1 &0 &0 &0\\ 0 &1 &0 &\!-1\\ 0 &0 &1 &2
\end{smallmatrix}\!\right)\,$,
so $\rank(P_{1,2})=3>N-1$.
So, by Lemma~\ref{L:rank}.i,
the index pair $[1,2]$ does not correspond to an asymptotic 
line soliton as $y\to\infty$.
(In fact, using Lemma~\ref{A:dominantphase} it can be verified that
$\theta_{3,5,6}$ is the only dominant phase combination along the line
$\theta_1=\theta_2$ as $y\to\infty$.)\,\ 
We then take $j=3$: in this case we have
$P_{1,3}= 
\left(\!\begin{smallmatrix} 
1 &0 &0\\ 0 &0 &\!-1\\ 0 &1 &2
\end{smallmatrix}\!\right)\,$,
with $\rank(P_{1,3})=2=:r$ and 
$\rank(P_{1,3}|A[1])= \rank(P_{1,3}|A[3])=
\rank(P_{1,3}|A[1,3])= 3= r+1$.
So the rank conditions in Lemma~\ref{L:rank}.i are satisfied.
Therefore the index pair $[1,3]$ corresponds to an asymptotic line soliton
as $y\to\infty$.
Moreover, by considering $j=4,5,6$ one can easily check that the 
rank conditions are no longer satisfied.
Thus $[1,3]$ is the \textit{unique} asymptotic line soliton associated
with the pivot index $e_1=1$ as $y\to\infty$, in agreement with 
Theorem~\ref{A:T1}.
Let us now consider the second pivot column, $e_2=4$.
In this case we find that the rank conditions are only satisfied when 
$j=5$, since 
$P_{4,5}= 
\left(\!\begin{smallmatrix}
1 &1 &1 &0\\ 0 &0 &0 &\!-1\\ 0 &0 &0 &2
\end{smallmatrix}\!\right)\,$,
with $\rank(P_{4,5})=2=:r$ and 
$\rank(P_{4,5}|A[4])= \rank(P_{4,5}|A[5])=
\rank(P_{4,5}|A[4,5])= 3= r+1$.
Therefore, the index pair $[4,5]$ corresponds to an asymptotic line soliton
as $y\to\infty$.
Finally, since $e_3=5$ and since we know from Theorem~\ref{A:T1},
that $j>e_3$, we immediately find that the third asymptotic
line soliton as $y\to\infty$ is given by the index pair $[5,6]$.
From Corollary~\ref{C:transition}, the phase transition diagram 
as $y\to\infty$ is given by
\begin{equation*}
R_{1,4,5} \mapto56 R_{1,4,6} \mapto45 R_{1,5,6} \mapto13 R_{3,5,6}\,.
\end{equation*}

The non-pivot columns of the coefficient matrix~$A$ are labeled by the indices
$g_1=2$ $g_2=3$ and $g_3=6$.
For $g_1=2$, the only possible value of $i<j$ is $i=1$.
In this case $Q_{1,2}=\emptyset$, so $\rank(Q_{1,2})=0$ and
$\rank(Q_{1,2}|A[1])= \rank(Q_{1,2}|A[2])= 
\rank(Q_{1,2}|A[1,2])= 1$.
Thus the pair $[1,2]$ identifies an asymptotic line soliton as $y\to-\infty$.
For $g_2=3$ we consider $i=2,1$:
when $i=2$, the rank conditions in Lemma~\ref{L:rank}.ii are satisfied,
leading to the asymptotic line soliton $[2,3]$ as $y\to-\infty$.
We can check that the soliton associated with the non-pivot column~$g_2=3$ 
is unique 
by considering $i=1$ and verifying that the rank conditions are not satisfied.
Similarly, it is easy to verify that for $g_3=6$ the index pair $[4,6]$
uniquely identifies the asymptotic line soliton as $y\to-\infty$.
The phase transition diagram as $y\to-\infty$ reads as follows:
\begin{equation*}
R_{1,4,5} \mapto12 R_{2,4,5} \mapto23 R_{3,4,5} \mapto46 R_{3,5,6}\,.
\end{equation*}

Summarizing, there are $N_+=3$ asymptotic line solitons as $y\to\infty$,
each associated with one of the pivots $e_1=1$, $e_2=4$ and $e_3=5$,
and indentified by the index pairs $[1,3]$, $[4,5]$ and $[5,6]$,
and there are $N_-=3$ asymptotic line solitons as $y\to-\infty$,
each associated with one of the non-pivot columns $g_1=2$, $g_2=3$ and $g_3=6$
and identified by the index pairs $[1,2]$, $[2,3]$ and $[4,6]$.
A snapshot of the solution at $t=-20$ is shown in Fig.~\ref{f:x3s}b.
\label{x:inelastic3s}
\end{example}

Examples~\ref{x:3to2} and~\ref{x:inelastic3s} illustrate the fact that, 
starting from any given coefficient matrix~$A$ in RREF, 
the asymptotic line solitons as $y\to\pm\infty$ can be identified 
in an algorithmic way by applying Theorem~\ref{A:T1} 
together with the rank conditions in Lemma~\ref{L:rank}.

\endgroup

\section{Further examples}
\label{s:examples}

In this section we further illustrate the asymptotic results 
derived in sections~\ref{s:general} and~\ref{s:asymptotics}
by discussing a variety of solutions of KPII generated by 
the tau-function~\eqref{e:tau} with different choices of 
coefficient matrices.

\begin{figure}[t!]
\newdimen\figwd
\newdimen\tboxht 
\newdimen\tboxwd
\figwd 0.375\textwidth
\tboxht 0.315\textwidth
\tboxwd -3.85em
\kern-\medskipamount
\centerline{\qquad\raise\tboxht\hbox{(a)}\hglue\tboxwd
\includegraphics[width=\figwd]{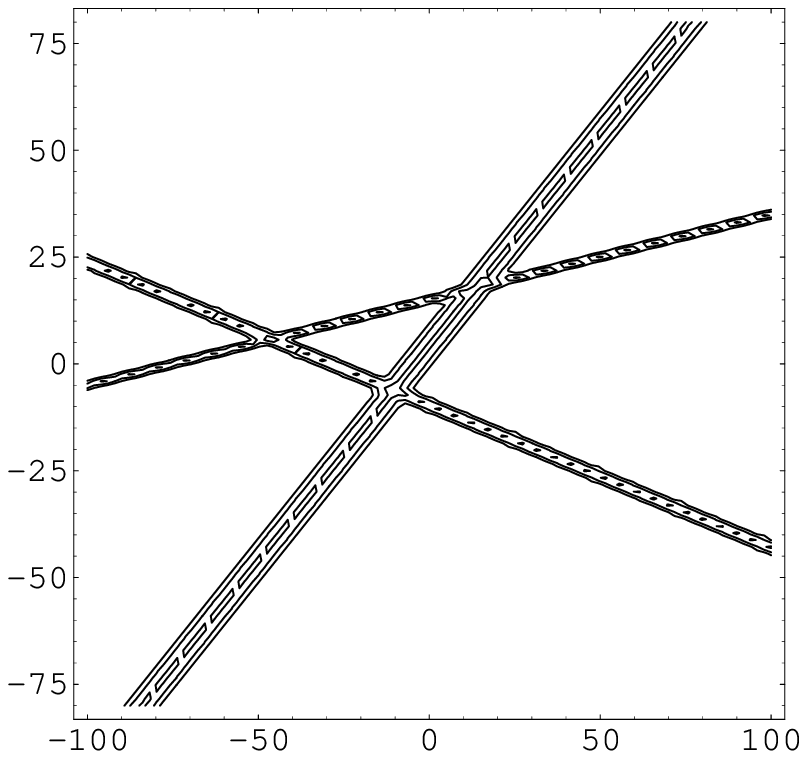}\qquad\qquad
\raise6ex\hbox{(b)}\hglue\tboxwd\raise0.5ex
\hbox{\includegraphics[width=0.975\figwd]{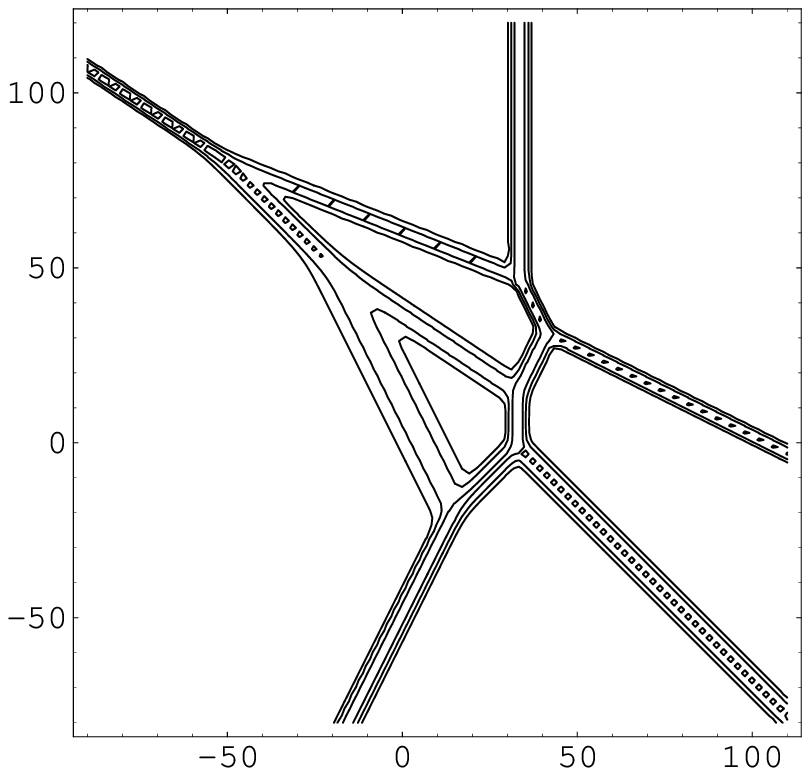}}\quad}
\smallskip
\centerline{\kern-1em\qquad\raise\tboxht\hbox{(c)}\hglue\tboxwd
\raise-0.2ex\hbox{\includegraphics[width=\figwd]{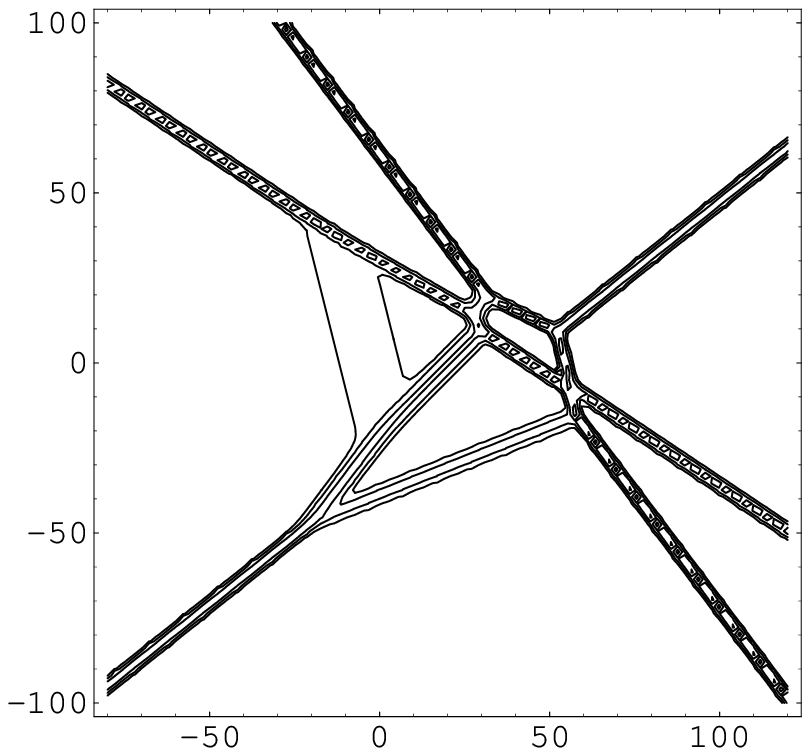}}\qquad\qquad\kern0.9em
\raise6ex\hbox{(d)}\hglue\tboxwd\kern-0.3em
\includegraphics[width=0.985\figwd]{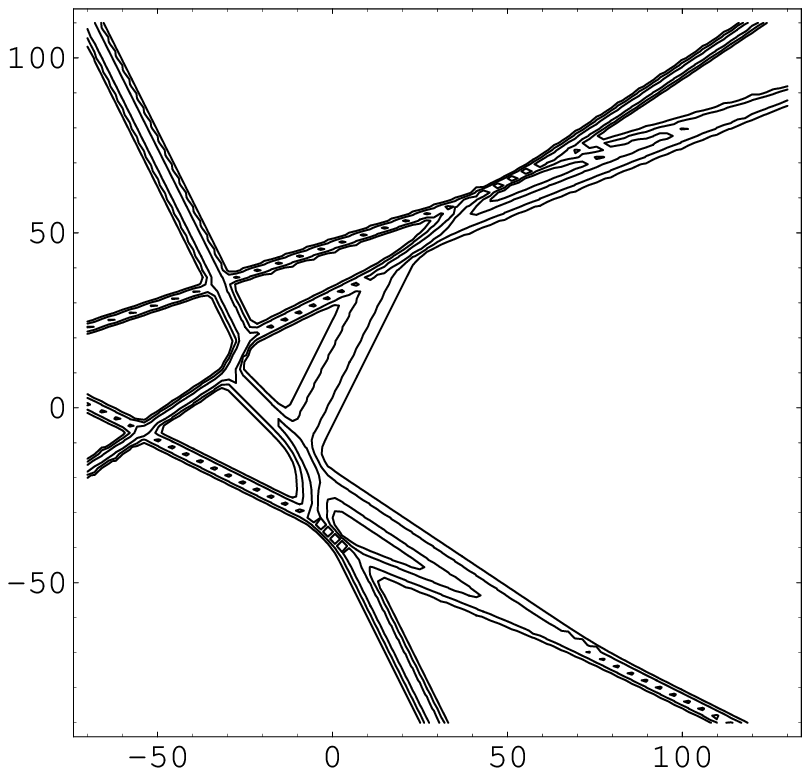}}
\smallskip
\centerline{\kern1.5em\raise\tboxht\hbox{(e)}\hglue\tboxwd
\raise0.3ex\hbox{\includegraphics[width=1.025\figwd]{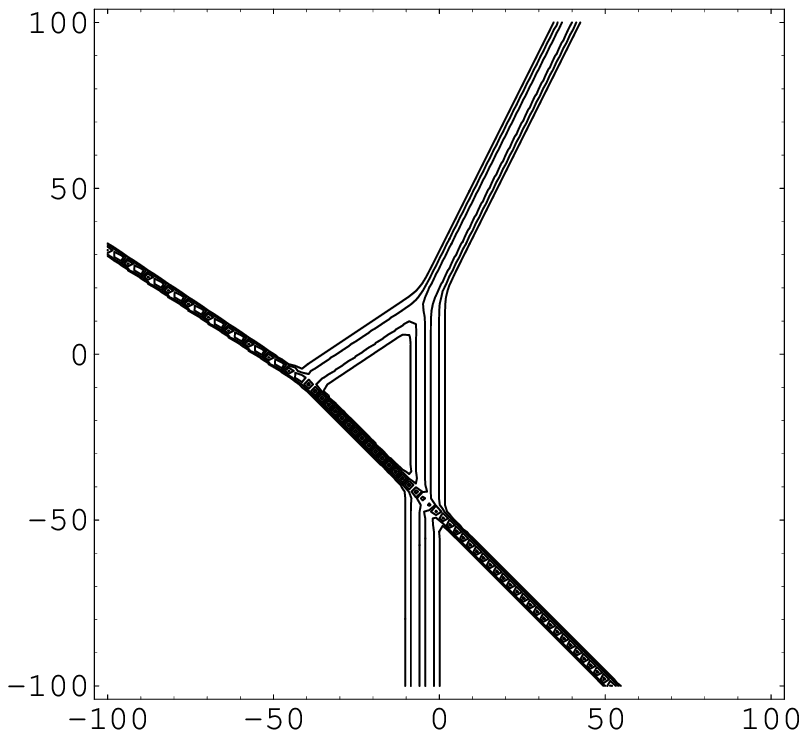}}\qquad\qquad
\raise\tboxht\hbox{(f)}\kern-0.0em\hglue\tboxwd
\includegraphics[width=1.025\figwd]{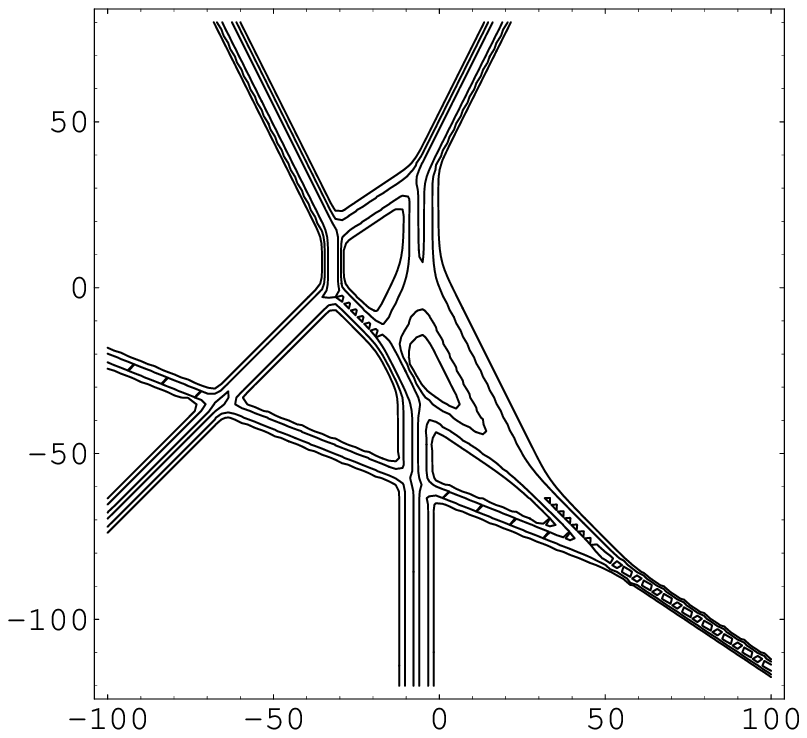}}
\caption{Line-soliton solutions of KPII:
(a)~an ordinary 3-soliton solution
with $(k_1,\dots,k_6)= (-3,-2,0,1,\frac32,2)$ at $t=4$;
(b)~a fully resonant (3,2)-soliton solution
with $(k_1,\dots,k_5)= (-1, 0, \frac12, 1, \frac32)$ at $t=-32$;
(c)~an elastic, partially resonant 3-soliton solution 
with~$A$ given by Eq.~\eqref{e:A3partialresonance}
and $(k_1,\dots,k_6)= (-\frac32, -1, 0, \frac14, \frac32, \frac74)$ at $t=-20$;
(d)~an elastic, partially resonant 4-soliton solution 
with~$A$ given by Eq.~\eqref{e:A4partial}
and $(k_1,\dots,k_8)=(-2,-\frac32,-1,-\frac12,0,\frac12,1,\frac32)$ at $t=20$;
(e)~an inelastic 2-soliton solution 
with~$A$ given by Eq.~\eqref{e:A2inelastic} 
and $(k_1,\dots,k_4)=(-1,-\frac12,\frac12,2)$ at $t=16$;
(f)~an inelastic 3-soliton solution
with~$A$ given by Eq.~\eqref{e:A3inelastic}
and $(k_1,\dots,k_6)=(-1,-\frac12,0,\frac12,1,\frac32)$ at $t=32$.}
\label{f:kps}
\kern-2\medskipamount
\end{figure}

\paragraph{Ordinary $N$-soliton solutions.}
These are constructed by taking $M=2N$ and choosing the functions 
$\{f_n\}_{n=1}^N$ in Eq.~\eqref{e:f} as (e.g., see Refs.~\cite{PLA95p1,MatveevSalle})
\begin{equation}
f_n(x,y,t)= e^{\theta_{2n-1}}+e^{\theta_{2n}}\,, \quad n=1,\dots,N\,.
\label{e:ordinaryNsoliton}
\end{equation}
The corresponding coefficient matrix is thus given by
\begin{equation*}
A =
  \begin{pmatrix}
     1 & 1 & 0 & 0 & \cdots & 0 & 0 \\
     0 & 0 & 1 & 1 & \cdots & 0 & 0 \\
     \vdots & \vdots & \vdots &\vdots & \vdots & \vdots & \vdots\\
     0 & 0 & 0 & 0 & \cdots & 1 & 1 \\
  \end{pmatrix}\,,
\label{e:Aordinary}
\end{equation*}
with $N$ pairs of identical columns at positions $\{2n-1,2n\},\, n=1,\dots,N$. 
There are only $2^N$ non-zero minors of $A$, which are
given by $A(m_1,m_2, \ldots, m_N)=1$ where, for each $n=1,\dots,N$,
either $m_n = 2n-1$ or $m_n=2n$.
The asymptotic analysis presented in the previous section 
allows one to identify these solutions as a subclass of 
elastic $N$-soliton solutions.
More precisely, the $N$ solitons are identified by the index pairs 
$[2n-1,2n]$ for $n=1,\dots,N$, where $i_n=2n-1$ and $j_n=2n$ label
respectively the pivot and non-pivot columns of~$A$.
Therefore their amplitudes and directions are given by
$a_n= k_{2n}-k_{2n-1}$ and $c_n= k_{2n-1}+k_{2n}$.
Moreover, the dominant pair of phase combinations for the $n$-th soliton 
as $y\to\infty$ is given by $\theta_{1,3,\dots,2n-1,2n+2,2n+4,\dots,2N}$ and
$\theta_{1,3,\dots,2n-3,2n,2n+2,\dots,2N}$,
while the dominant phase combinations for the same soliton as $y\to-\infty$ by 
$\theta_{2,4,\dots,2n,2n+1,2n+3,\dots,2N-1}$ and
$\theta_{2,4,\dots,2n-2,2n-1,2n+1,\dots,2N-1}$.
Apart from the position shift of each soliton,
the interaction gives rise to a pattern of $N$ intersecting
lines in the $xy$-plane, as shown in Fig.~\ref{f:kps}a.

\paragraph{Solutions of KPII which also satisfy the finite Toda lattice hierarchy.}

Another class of $(N_-,N_+)$-soliton solutions of KPII is given
by the following the choice of functions $\{f_n\}_{n=1}^N$ in Eq.~\eqref{e:f}:
\begin{gather}
f_n = f^{(n-1)} \,\quad n=1,\ldots,N \,.
\label{e:resonantNsoliton}
\end{gather}
In addition to generating solutions of KPII, 
the set of tau-functions $\tau_{N,M}$ for $N=1,\dots,M$ 
also satisfy the Pl\"{u}cker relations for the
finite Toda lattice hierarchy~\cite{jphysa36p10519}.
Choosing $f(x,y,t)=\sum\nolimits_{m=1}^M e^{\theta_m}\,$
then yields the following coefficient matrix:
\begin{equation}
A =
  \begin{pmatrix}
     1 & 1 & \cdots & 1  \\
     k_1 & k_2 & \cdots & k_M \\
     \vdots & \vdots & \ddots & \vdots \\
     k_1^{N-1} & k_2^{N-1} & \cdots & k_M^{N-1}
  \end{pmatrix}\,.
\label{e:AToda}
\end{equation}
Note that $A$ in Eq.~\eqref{e:AToda} is not in RREF yet,
and coincides with the matrix~$K$ in Lemma~\ref{L:binetcauchy}.
Here the pivot columns are labeled by indices $1,\dots,N$;
all the $\binom MN$ minors of~$A$ are non-zero,
and coincide with the Van~der~Monde determinants~\eqref{e:vandermonde}.
This class of solutions was studied in Ref.~\cite{jphysa36p10519},
where it was shown that the $N$ asymptotic line solitons as $y\to\infty$ 
are identified by the index pairs~$[n,n+M-N]$ for $n=1,\dots,N$,
while the $M{-}N$ asymptotic line solitons as $y\to-\infty$
are identified by the index pairs~$[n,n+N]$ for $n=1,\dots,M-N$.
These pairings can also be easily verified using Theorem~\ref{A:T1}.
The dominant pair of phase combinations for the $n$-th soliton as $y\to\infty$
is given by $\theta_{1,\dots,n,M-N+n+1,\dots,M}$ and 
$\theta_{1,\dots,n-1,M-N+n,\dots,M}$, 
while the dominant pair of phase combinations for the $n$-th soliton 
as $y\to-\infty$ by $\theta_{n,\dots,N+n-1}$ and $\theta_{n+1,\dots,N+n}$.
The solution displays phenomena of soliton resonance 
and web structure (e.g., see Fig.~\ref{f:kps}b).
More precisely, the interaction of the asymptotic line solitons results 
in a pattern with 
$(2N_-{-}1)N_+$~interaction vertices, 
$(3N_-{-}2)N_+$~intermediate interaction segments 
and $(N_-{-}1)(N_+{-}1)$~``holes'' in the $xy$-plane.
Each of the intermediate interaction segment can be 
effectively regarded as a line soliton since it satisfies 
the dispersion relation~\eqref{e:dispersionrelation}.
Furthermore, all of the asymptotic and intermediate line solitons 
interact via a collection of fundamental resonances:
a~\textit{fundamental resonance}, also called a Y-junction,
is a travelling-wave solution of KPII describing
an intersection of three line solitons 
whose wavenumbers~$\@k_a$ and frequencies~$\omega_a$ ($a=1,2,3$)
satisfy the three-wave resonance 
conditions~\cite{JFM1977v79p171,NewellRedekopp} 
\begin{equation}
\@k_1 + \@k_2 = \@k_3\,,\qquad
\omega_1 + \omega_2 = \omega_3\,.
\label{e:resonance}
\end{equation}
Such a solution is shown in Fig.~\ref{f:dominantphases}a.

\paragraph{Elastic $N$-soliton solutions.}

As mentioned in sections~\ref{s:introduction} and~\ref{s:asymptotics},
elastic $N$-soliton solutions are those for which 
the sets of incoming and outgoing asymptotic line solitons are the same.
In this case we necessarily have $M=2N$.
Ordinary $N$-soliton solutions and solutions of KPII which also 
satisfy the finite Toda lattice hierarchy with $M=2N$ are two 
special classes of elastic $N$-soliton solutions.
However, a large variety of other elastic $N$-soliton solutions 
do also exist.
For example, Fig.~\ref{f:kps}c shows an elastic
3-soliton solution generated by the coefficient matrix:
\begin{equation}
A= \begin{pmatrix}
    1 &0 &0 &1 &1 &1\\
    0 &1 &0 &\!-2 &\!-2 &\!-1\\
    0 &0 &1 &2 &1 &0
\end{pmatrix}\,.
\label{e:A3partialresonance}
\end{equation}
In this case the pivot columns are labeled by indices 1, 2 and 3.
So, from Lemma~\ref{L:pivot} we know that 
the asymptotic line solitons as $y\to\infty$
will be identified by index pairs $[1,j_1]$, $[2,j_2]$ and $[3,j_3]$,
while those as $y\to-\infty$ by index pairs $[i_1,4]$, $[i_2,5]$ and $[i_3,6]$,
for some value of $i_1,\dots,i_3$ and $j_1,\dots,j_3$.
Indeed,
use of the asymptotic techniques developed in section~\ref{s:asymptotics}
allows one to conclude that both the incoming and the outgoing 
asymptotic line solitons 
are given by the \textit{same} index pairs $[1,4]$, $[2,6]$ and $[3,5]$.
The soliton interactions in this case are \textit{partially resonant},
in the sense that 
the pairwise interaction among solitons $[1,4]$~and~$[2,6]$ 
and that among solitons $[1,4]$~and~$[3,5]$ are both resonant,
but the pairwise interaction among solitons $[2,6]$~and~$[3,5]$
is non-resonant.
Similarly, Fig.~\ref{f:kps}d shows an elastic,
partially resonant 4-soliton solution generated by the coefficient matrix
\begin{equation}
A= \begin{pmatrix}
    1 &0 &\!-1 &0 &1 &0 &\!-1 &\!-2\\
    0 &1 &2 &0 &\!-1 &0 &1 &2\\
    0 &0 &0 &1 &2 &0 &-1 &\!-2\\
    0 &0 &0 &0 &0 &1 &2 &3
\end{pmatrix}\,.
\label{e:A4partial}
\end{equation}
In this case the pivot columns are labeled by the indices 1, 2, 4 and~6
and the non-pivot columns by the indices 3, 5, 7 and~8.
The asymptotic line solitons as $y\to\pm\infty$ are identified by 
the index pairs $[1,3]$, $[2,5]$, $[4,7]$ and $[6,8]$.
As can be seen from Fig.~\ref{f:kps}f, 
the pairwise interaction of solitons $[1,3]$~and~$[2,5]$,
solitons $[2,5]$~and~$[4,7]$, and $[4,7]$~and~$[6,8]$ are resonant, 
but all other pairwise interactions
(e.g., the pairwise interactions between 
solitons $[1,3]$~and~$[4,7]$, $[1,3]$~and~$[6,8]$, $[2,5]$~and~$[6,8]$) 
are non-resonant.
It should be clear from these examples that a large variety of elastic
$N$-soliton solutions with resonant, partially resonant and non-resonant
interactions is possible.
The properties of elastic $N$-soliton solutions are studied in 
detail in Refs.~\cite{GBSCYK,Kodama}.

\paragraph{Inelastic $N$-soliton solutions.}
$N$-soliton solutions that are not elastic are called inelastic.
We have already seen such solutions 
in Examples~\ref{x:3to2} and~\ref{x:inelastic3s}
(cf.\ Figs.~\ref{f:x3s}a,b) of section~\ref{s:asymptotics}.
As a further example, Fig.~\ref{f:kps}e shows an inelastic 2-soliton solution 
generated by the coefficient matrix
\begin{equation}
A= \begin{pmatrix}
1 &0 &0 &\!-1\\[0.2ex]
0 &1 &1 &1
\end{pmatrix}\,. 
\label{e:A2inelastic}
\end{equation}
In this case the pivot columns are labeled by indices 1 and 2;
the asymptotic line solitons as $y\to-\infty$ are identified 
by the index pairs $[1,4]$ and $[2,3]$,
while those as $y\to\infty$ by the index pairs $[1,3]$ and $[2,4]$.
Notice that the outgoing solitons interact resonantly (via two Y-junctions), 
while the incoming soliton pair interact non-resonantly.
This is in contrast with an elastic $2$-soliton solution, 
where both incoming and outgoing pairs of solitons exhibit 
the same kind of interaction.
Similarly, Fig.~\ref{f:kps}f shows inelastic 3-soliton solution
generated by the coefficient matrix
\begin{gather}
A= \begin{pmatrix}
1 &0 &\!-1\! &\!-1\! &0 &2\\[0.2ex]
0 &1 &2    &1    &0 &\!-1\!\\[0.2ex]
0 &0 &0    &0    &1 &1
\end{pmatrix}\,.
\label{e:A3inelastic}
\end{gather}
Here the pivot columns are labeled by indices 1, 2 and 5;
the asymptotic line solitons as $y\to\infty$ are identified 
by the index pairs $[1,3]$, $[2,5]$ and $[5,6]$,
while those as $y\to-\infty$ by the index pairs $[1,3]$, $[2,4]$ and $[3,6]$.
Finally, in the generic case one has $M\ne 2N$, and
the numbers of asymptotic line solitons as $y\to\pm\infty$ are different,
as in the solutions shown in Figs.~\ref{f:x3s}a and~\ref{f:kps}b.

We should point out that one-soliton solutions, 
ordinary two-soliton solutions and fundamental resonances
have the property that their time evolution is just an overall 
translation of a fixed spatial pattern.
The same property does not hold, however, for all the other solutions 
presented in this work.  That is,
the interaction patterns formed by these line solitons, and the
relative positions of the interaction vertices in the 
$xy$-plane are in general time-dependent.

\section{Conclusions}
\label{s:conclusions}

In this article we have studied a class of line-soliton solutions 
of the Kadomtsev-Petviashvili II equation by expressing the tau-function as 
the Wronskian of $N$ linearly independent combinations of $M$ exponentials.
From the asymptotics of the tau-function as $y\to\pm\infty$
we showed that each of these solutions of KPII is composed of 
asymptotic line solitons which are defined by the transition 
between two dominant phase combinations with $N{-}1$ common phases.
Moreover, the number, amplitudes and directions of the
asymptotic line solitons are invariant in time.
We also derived an algorithmic method to identify these 
asymptotic line solitons in a given solution
by examining the $N\times M$ coefficient matrix~$A$
associated with the corresponding tau-function.
In particular, we proved that 
every $N\times M$, irreducible coefficient matrix~$A$ produces an 
$(N_-,N_+)$-soliton solution of KPII in which there are 
$N_+=N$ asymptotic line solitons as $y\to\infty$,
labeled by the pivot columns of~$A$,
and $N_-=M{-}N$ asymptotic line solitons as $y\to-\infty$,
labeled by the non-pivot columns of~$A$.
Such solutions exhibit a rich variety of time-dependent spatial patterns 
which include resonant soliton interactions and web structure.
Finally, we discussed a number of examples of such 
$(N_-,N_+)$-soliton solutions in order to illustrate the above results.

It is remarkable that the KPII equation possesses 
such a rich structure of line-soliton solutions which are
generated by a simple form of the tau-function.
In this work we have primarily focused on the asymptotic behavior
of the solutions as $y\to\pm\infty$, but not on their interactions
in the $xy$-plane. A full characterization of the interaction 
patterns of the general $(N_-,N_+)$-soliton solutions is 
an important open problem, which is left for further study.
Nonetheless, we believe that our results will provide a key step
toward that endeavor. 
Solutions exhibiting phenomena of soliton resonance and web structure 
have been found for several other (2+1)-dimensional integrable systems,
and those solutions can also be described by direct algebraic methods
similar to the ones used here.
Therefore we expect that the results {\bf presented} in this work 
will also be useful to study solitonic solutions 
in a variety of other (2+1)-dimensional integrable systems.

\section*{Acknowledgements}

It is a pleasure to thank M.~J.~Ablowitz and Y.~Kodama for 
many insightful discussions.


\appendix
\section*{Appendix}
\renewcommand\theequation{\Alph{section}.\arabic{equation}}
\renewcommand\thesubsection{\Alph{section}.\arabic{subsection}}
\catcode`\@ 11
\catcode`\@ 12
\setcounter{section}1
\def\sum{\mathop{\textstyle\truesum}\nolimits}
\def\span{\mathop{\rm span}\nolimits}

\subsection{Proof of Theorem \ref{T:transition}}

To prove part~(i) of Theorem \ref{T:transition},
it is sufficient to show that, along each line~$L_c$,
the sign of the inequalities among the
phase combinations in Definition~\ref{D:dominantphase}
remain unchanged in time as $y\to\pm\infty$.
For this purpose, note that the sign of
$\theta_{m_1,\ldots,m_N} - \theta_{m'_1,\ldots,m'_N}$
in Eq.~\eqref{e:dominantphase} is determined by the coefficient of $y$
on the right-hand side as $y\to\pm\infty$ and for finite~$\xi$ and~$t$,
if this coefficient is non-zero.
For generic values of the phase parameters $k_1,\dots,k_M$
this coefficient is indeed non-vanishing, and its sign
depends only on the direction~$c$ of the line~$L_c$.
Consequently, the dominant phase combinations
asymptotically as~$y\to\pm\infty$
are determined only by the constant~$c$ for finite time.

Part~(ii) of the theorem is proved by showing that the only possible
phase transitions are those in which a single phase, say $\theta_m$
changes to $\theta_{m'}$
between the two dominant phase combinations across adjacent regions,
and that no other type of transitions can occur.
We first prove that single-phase transitions are allowed;
then we show that no other type of transitions are allowed.
In the following, we will assume $t$ to be finite so that
the dominant phase combinations remain invariant, according to part (i).
Suppose that $\theta_{m_1,\dots,m_N}$ is the dominant phase combination
in a region~$R$ asymptotically for large values of~$|y|$.
Since~$R$ is a proper subset of~$\Real^3$, it \textit{must} have a boundary,
across which a transition will take place from
$\theta_{m_1,\dots,m_N}$ to some other dominant phase combination.
Since $\theta_{m_1,\dots,m_N}$ is dominant,
$A(m_1,\dots,m_N)\ne 0$ according to Definition~\ref{D:dominantphase}.
Therefore, the columns $A[m_1],\dots,A[m_N]$ of the coefficient matrix
form a basis of $\Real^N$, and for all $j \notin \{ m_1, \ldots, m_N \}$
we have that $A[j]$ is in the span of $A[m_1],\dots,A[m_N]$.
Thus there exists at least one column~$A[m_s]$ such that
the coefficient of $A[m_s]$ in the expansion of $A[j]$ is non-zero.
We then have $A(m_1,\dots,m_{s-1},j,m_{s+1},\dots,m_N)\ne 0$,
implying that the phase combination
$\theta_{m_1,\dots,m_{s-1},j,m_{s+1},\dots,m_N}$
is actually present in the tau-function.
Then, for \textit{any} $j \notin \{m_1,\ldots,m_N\}$
it is possible to have a single-phase transition from~$R$
to the adjacent region~$R'$ across the line $\theta_{m_s} = \theta_j$,
since the sign of $\theta_{m_s}-\theta_j$ changes across this line,
implying that $\theta_{m_1,\dots,m_{s-1},j,m_{s+1},\dots,m_N}$
is larger than $\theta_{m_1,\dots,m_N}$ in~$R'$.

We next prove that no other type of transitions can occur
apart from single-phase transitions;
we do so by \textit{reductio ad absurdum}.
Suppose that at least two phases~$\theta_{m_1}, \theta_{m_2}$
from the dominant phase combination~$\theta_{m_1,\dots,m_N}$
in a region $R$ are replaced with phases~$\theta_{m'_1}, \theta_{m'_2}$
during the transition from~$R$ to an adjacent region~$R'$.
This transition occurs along the common boundary
of $R$ and $R'$, which is given by line
$L:(\theta_{m_1}+\theta_{m_2})-(\theta_{m'_1}+\theta_{m'_2}) = 0$,
Thus, along~$L$, the differences
$\theta_{m_1}-\theta_{m'_1}$ and $\theta_{m_2}-\theta_{m'_2}$
(or, equivalently, the differences
$\theta_{m_1}-\theta_{m'_2}$ and $\theta_{m_2}-\theta_{m'_1}$)
must have opposite signs or be both zero.

If both differences are zero along~$L$, the lines
$\theta_{m_1}=\theta_{m_1'}$ and $\theta_{m_2}=\theta_{m_2'}$
(or, equivalently, the lines
$\theta_{m_1}=\theta_{m_2'}$ and $\theta_{m_1'}=\theta_{m_2}$)
must both coincide with the line~$L$ in the $xy$-plane.
This is possible only at a given instant of time \textit{and}
if the directions of the two lines are the same,
i.e., if $k_{m_1}+k_{m_1'}=k_{m_2}+k_{m_2'}$
(or, equivalently, $k_{m_1}+k_{m_2'}=k_{m_1'}+k_{m_2}$).
So for generic values of the phase parameters, or for generic
values of time, this exceptional case can be excluded.
Hence, we assume that
$\theta_{m_1}-\theta_{m_1'}$ and $\theta_{m_2}-\theta_{m_2'}$
are of opposite signs.
Note however that
$\theta_{m_1}-\theta_{m'_1}=
  \theta_{m_1,\dots,m_N} - \theta_{m'_1,m_2,\dots,m_N}$
and $\theta_{m_2}-\theta_{m'_2}=
  \theta_{m_1,\dots,m_N} - \theta_{m_1,m'_2,m_3\dots,m_N}$.
Moreover,
both of these phase differences must be \textit{positive}
in the interior of~$R$ if the minors
$A(m'_1,m_2,\dots,m_N)$ and $A(m_1,m'_2,m_3\dots,m_N)$ are non-zero,
since $\theta_{m_1,\dots,m_N}$ is the dominant phase in~$R$.
Hence, we must conclude that
$\theta_{m_1}-\theta_{m_1'}$ and $\theta_{m_2}-\theta_{m_2'}$
cannot have opposite signs unless one or both of the
phase combinations $\theta_{m'_1,m_2,\dots,m_N}$ and 
$\theta_{m_1,m'_2,m_3\dots,m_N}$ is absent from the tau-function. 
This requires that either
$A(m'_1,m_2,\dots,m_N)$ or $A(m_1,m'_2,m_3\dots,m_N)$ must be zero.
A similar argument applied to the the phase differences
$\theta_{m_1}-\theta_{m'_2}$ and $\theta_{m_2}-\theta_{m'_1}$
leads to the conclusion that one or both of the minors
$A(m'_2,m_2,\dots,m_N)$ and $A(m_1,m'_1,m_3\dots,m_N)$ must vanish.
However, from the Pl\"ucker relations among the $N\times N$ minors of~$A$
we have
\begin{multline}
A(m_1,m_2\dots,m_N)A(m'_1,m'_2,\dots,m_N) =
\\
A(m_1,m'_2,m_3,\dots,m_N)
A(m'_1,m_2,\dots,m_N) - A(m_1,m'_1,m_3\dots,m_N)A(m'_2,m_2,\dots,m_N)\,.
\end{multline}
Then it follows that either
$A(m_1,\dots,m_N)=0$ or $A(m'_1,m'_2,m_3,\dots,m_N)=0$.
But this is impossible since by assumption both minors on the
left-hand-side are associated with dominant phase combinations.
Thus, they are both non-zero. Hence we have a reached a contradiction
which implies that as~$y\to\pm\infty$, phase transitions where more than 
one phase changes simultaneously across adjacent dominant phase 
regions, are impossible.

\subsection{Proof of Theorem \ref{A:T1}}

First we need to establish the following Lemma that
will be useful in proving the theorem.
\begin{lemma}
If $P_{ij}$ is the submatrix defined in Eq.~\eqref{e:PQ} and
$e_n$ labels the $n$-th pivot column of an irreducible coefficient matrix~$A$,
then
$N-1 \le \rank(P_{e_ne_{n}+1}) \le N,\,\forall n=1,\ldots,N$.
\label{A:rD}
\end{lemma}
\begin{proof}
Recall that the pivot indices are ordered as $1=e_1<e_2<\ldots<e_N<M$
for an irreducible matrix $A$. Then it follows from 
Definition~\ref{A:D1}.ii that, corresponding to each pivot column 
$A[e_n]$ of an irreducible matrix~$A$,
there exists at least one non-pivot column $A[j_*]$, with $j_*>e_n$,
that has a non-zero entry in its $n$-th row.
Hence we have $A(e_1,\ldots,e_{n-1},j_*,e_{n+1},\ldots,e_N)\ne0$.
This implies that the matrix
$A[1,\ldots,e_n-1,e_n+1,\ldots,M] = (P_{e_ne_{n}+1}|A[e_n+1])$
which contains the columns 
$A[e_1], \ldots, A[e_{n-1}], A[j*], A[e_{n+1}], \ldots, A[e_N]$,
has rank~$N$. Thus, the rank of $P_{e_ne_{n}+1}$ is {\em at least} $N-1$,
and this yields the desired result.
\end{proof}

We are now ready to prove Theorem~\ref{A:T1}.
We prove part~(i) here; the proof of part~(ii) follows similar steps.
The proof is divided in two parts.
First we show that for each pivot index~$e_n,\,n=1,\dots,N$,
there exists an index~$j_n>e_n$ with the necessary and sufficient
properties for $[e_n,j_n]$ to identify an asymptotic line soliton
as $y\to\infty$;
then we prove that such a~$j_n$ is unique.
\par
\textit{Existence.}~
The proof is constructive.
For each pivot index~$e_n$, and for any $j>e_n$,
we consider the rank of the matrix
$P_{e_n,j}= A[1,2,\dots,e_n-1,j+1,\dots,M]$
starting from $j=e_n+1$.
When $j=e_n+1$ we have $P_{e_n,j}=P_{e_n,e_n+1}$, and therefore
$N-1\le\rank(P_{e_n,e_n+1})\le N$ from Lemma~\ref{A:rD}.
If $\rank(P_{e_n,e_n+1})=N$, then Lemma~\ref{L:rank}.i
implies that the pair $[e_n,e_n+1]$
\textit{does not} identify an asymptotic line soliton as $y\to\infty$.
In this case, we increment the value of $j$ successively from $e_n+1$,
until
\unskip\footnote{Note that a value of $j$ such that $\rank(P_{e_n,j})=N-1$
always exists, since for $j=M$ we have $P_{e_n,M}=A[1,\dots,e_n-1]$
whose rank is $n-1$, since $A$ is in RREF.}
$\rank(P_{e_n,j})$ decreases from $N$ to $N-1$.
Suppose $j=j_*$ is the smallest index such that
$\rank(P_{e_n,j_*})=N-1$ and $\rank(P_{e_n,j_*}|A[j_*])=N$.
We next check the rank of $\rank(P_{e_n,j_*}|A[e_n])$.
Since $\rank(P_{e_n,j_*})=N-1$,
two cases are possible: either
(a)~$\rank(P_{e_n,j_*}|A[e_n])=N$ or
(b)~$\rank(P_{e_n,j_*}|A[e_n])=N-1$.
We discuss these two cases separately.

(a)~
Suppose that \,$\rank(P_{e_n,j_*}|A[e_n])=N$.
By construction we have \,$\rank(P_{e_n,j_*}|A[j_*])=N$\,,
and since $N=\rank(A)$ one also has
$\rank(P_{i_n,j_*}|A[e_n,j_*])=N$.
In this case we set $j_*=j_n$.
It follows from Lemma~\ref{L:rank} that the pair $[e_n,j_n]$ satisfies
the necessary rank conditions to identify
an asymptotic line soliton as $y\to\infty$.
Next we show that these rank conditions are also sufficient in order to
determine a pair of dominant phase combinations in the tau function
corresponding to the single-phase transition $e_n\to j_n$.
Since $\rank(P_{e_n,j_n})=N-1$, it is possible to choose $N-1$
linearly independent columns $A[p_1],\dots,A[p_{N-1}]$
from the matrix $P_{e_n,j_n}$ so that
for all choices of linearly independent columns
$A[l_1],\dots,A[l_{N-1}]\in P_{e_n,j_n}$ one has
\unskip\footnote{The existence of such a set is guaranteed because
part~(i) of the dominant phase condition~\ref{A:dominantphase}
implies that, as $y\to\infty$ in the $[e_n,j_n]$ direction,
the phases corresponding to the index set $P_{e_n,j_n}$
are ordered as $\theta_1>\theta_2>\dots>\theta_{e_n-1}$
and $\theta_{j_n+1}<\theta_{j_n+2}<\dots<\theta_M$.
Then, since $\rank(P_{e_n,j_n})=N-1$,
it is possible to select the top $N-1$ phases from the above two lists
so that the corresponding columns are linearly independent.}
$\theta_{p_1,\dots,p_{N-1}}\ge \theta_{l_1,\dots,l_{N-1}}$
as $y\to\infty$ along the transition line $L_{e_n,j_n}$.
Furthermore, since
$\rank(P_{e_n,j_n}|A[e_n]) = \rank(P_{e_n,j_n}|A[j_n])=N$,
the minors $A(e_n,p_1,\dots,p_r)$ and
$A(j_n,p_1,\dots,p_r)$ are both non-zero, and thus
$\theta_{e_n,p_1,\dots,p_{N-1}}$ and $\theta_{j_n,p_1,\dots,p_{N-1}}$
form a dominant pair of phase combinations as $y\to\infty$ along the
direction of $L_{e_n,j_n}$.

(b)~
Suppose that $\rank(P_{e_n,j_*}|A[e_n])=N-1$.
\unskip\footnote{Note that this is possible only for $n<N$,
because when $n=N$ the submatrix $P_{e_N,j}$ for any $j>e_N$
contains the pivot columns $A[e_1],\dots,A[e_{N-1}]$.
Hence, $\rank(P_{e_N,j})=N-1$ and $\rank(P_{e_N,j}|A[e_N]) = N$.
Consequently, $n=N$ always belongs to case~(a) above
and not to case~(b).}
Since $\rank(P_{e_n,j_*})=N-1$ by construction, this means that
$A[e_n]\in\span(P_{e_n,j_*})$.
However, since $A[e_n]$ is a pivot column,
it cannot be spanned by its preceding columns $A[1],\dots,A[e_n-1]$.
Hence the spanning set of~$A[e_n]$ from $P_{e_n,j_*}$
must contain at least one column from $A[j_*+1],\dots,A[M]$.
In this case we continue incrementing the value of $j$ starting from $j_*$
until the pivot column $A[e_n]$ is no longer in the span of the columns
of the resulting submatrix $P_{e_n,j}$.
Let $j_n$ be the smallest index such that $A[e_n]$ is spanned by the
columns of~$P_{e_n,j_n}|A[j_n]$ but \textit{not} by those of~$P_{e_n,j_n}$.
Then, by construction we have $\rank(P_{e_n,j_n})=:r<N-1$, and
$\rank(P_{e_n,j_n}|A[e_n])= \rank(P_{e_n,j_n}|A[j_n])=
\rank(P_{e_n,j_n}|A[e_n,j_n])=r+1$.
The rank conditions in Lemma~\ref{L:rank}.i are once again satisfied
for the index pair~$[e_n,j_n]$ thus found.
The sufficiency of these conditions can then be established by following
similar steps as in case~(a).
Namely, it is possible to choose a set of linearly independent vectors
$A[l_1],\dots,A[l_r]\in P_{e_n,j_n}$ and extend it to a basis
$\{A[e_n],A[l_1],\dots,A[l_r],A[m_1],\dots,A[m_s]\}$ of $\Real^N$,
where $A[m_1],\dots,A[m_s]\in Q_{e_n,j_n}$ and $r+s=N-1$.
We then have $A(e_n,l_1,\dots,l_r,m_1,\dots,m_s)\ne0$,
which also implies  $A(j_n,l_1,\dots,l_r,m_1,\dots,m_s)\ne0$
since $A[e_n]\in\span(P_{e_n,j_n}|A[j_n])$.
As in case~(a), we can now maximize the phase combinations
over all such sets $\{l_1,\dots,l_r\,m_1,\dots,m_s\}$,
and find a set of indices $\{p_1,\dots,p_r,q_1,\dots,q_s\}$
such that
$\theta_{e_n,p_1,\dots,p_r,q_1,\dots,q_s}$
and $\theta_{j_n,p_1,\dots,p_r,q_1,\dots,q_s}$
form a dominant pair of phase combinations as $y\to\infty$ along the
direction of $L_{e_n,j_n}$.
Summarizing, we have shown that for each pivot index $e_n$, $n=1,2,\dots,N$,
there exists at least one asymptotic line soliton $[e_n,j_n]$
with $j_n>e_n$ as $y\to\infty$.  Next we prove uniqueness.
                                                                                
\textit{Uniqueness.}~
Suppose that $[e_n,j_n]$ and $[e_n,j_n']$ are
two asymptotic line solitons identified by the same pivot index $e_n$
as $y\to\infty$.
Without loss of generality, assume that $j_n'>j_n$,
and consider the line soliton $[e_n,j_n']$.
Lemma~\ref{L:rank}.i implies that
$\rank(P_{e_n,j_n'}|A[j_n'])=\rank(P_{e_n,j_n}|A[e_n,j_n'])$.
Hence the pivot column $A[e_n]$ is spanned by the columns of the
submatrix~$(P_{e_n,j_n'}|A[j_n'])$.
But by assumption we have 
$(P_{e_n,j_n'}|A[j_n'])\subseteq P_{e_n,j_n}$, since $j_n'>j_n$.
Hence $A[e_n]$ is also spanned by the columns of $P_{e_n,j_n}$.
This however implies that $\rank(P_{e_n,j_n})=\rank(P_{e_n,j_n}|A[e_n])$,
which contradicts the necessary rank conditions in Lemma~\ref{L:rank}.i
for $[e_n,j_n]$ to identify an asymptotic line soliton as $y\to\infty$.
Therefore we must have $j_n=j_{n'}$. Thus, it is not possible to have
two distinct asymptotic line solitons
as $y\to\infty$ associated with the same pivot index~$e_n$.
Part~(i) of Theorem~\ref{A:T1} is now proved.

\subsection{Equivalence classes and duality of solutions}

In this appendix, we investigate the relationship between two classes of
KPII multi-soliton solutions with complementary sets of asymptotic
line solitons.
Note that the KPII equation~\eqref{e:KP}
is invariant under the inversion symmetry $(x,y,t) \to (-x,-y,-t)$.
As a result, if $u(x,y,t)$ is an $(M-N,N)$-soliton solution of KPII
with $M-N$ incoming and $N$ outgoing line solitons, then $u(-x,-y,-t)$ is
a $(N,M-N)$-soliton solution of KPII where the numbers of incoming and
outgoing line solitons are reversed. It follows from Theorem~\ref{A:T1}
that the solution $u(-x,-y,-t)$ must correspond to \textit{some}
tau-function $\tau_{M-N,M}(x,y,t)$ associated with an $M-N \times M$
coefficient matrix whose pivot and non-pivot columns uniquely identify
the asymptotic line solitons of $u(-x,-y,-t)$.
Before proceeding further, we introduce the notion of an equivalence class
which plays an important role in subsequent discussions.
Let $\Theta$ denote the set of all phase combinations
$\theta_{m_1,\dots,m_N}$ which appear with nonvanishing coefficients
in the tau-function~$\tau(x,y,t)$ of Eq.~\eqref{e:tauphases}.
\begin{definition}
(Equivalence class)~
Two tau-functions are defined
to be in the same equivalence class if (up to an overall exponential
phase factor) the set $\Theta$ is the same for both. The set of
$(N_-,N_+)$-soliton
solutions of KPII generated by an equivalence class of tau-functions
defines an equivalence class of solutions.
\label{D:equivalenceclass}
\end{definition}
It is clear from the above definition that tau-functions
in a given equivalence class can be viewed as positive-definite sums
of the \textit{same} exponential phase combinations but with
different sets of coefficients. They
are parametrized by the same set of phase parameters $k_1,\ldots,k_M$,
but the constants $\theta_{m0}$ in the phase $\theta_m$ are different.
Moreover, the irreducible coefficient matrices associated with
the tau-functions have exactly the same sets of vanishing and
non-vanishing minors, but the magnitudes of the non-vanishing minors
are different for different matrices.
The asymptotic line solitons of each solution in an
equivalence class arise from the {\em same} $i \to j$ single phase
transition, and are therefore labeled by the same
index pair $[i,j]$. Theorem~\ref{A:T1}
then implies that the coefficient matrices associated with
the tau-functions in the same equivalence class have identical
sets of pivot and non-pivot indices which identify respectively,
the asymptotic line solitons as $y \to \infty$ and as $y \to -\infty$.
Thus, solutions in the same equivalence class can differ only
in the position of each asymptotic line
solitons and in the location of each interaction vertex. As a result,
any $(N_-,N_+)$-soliton solution of KPII can be transformed into
any other solution in the same equivalence class by spatio-temporal
translations of the individual asymptotic line solitons.
We refer to the two tau-functions $\tau_{N,M}(x,y,t)$ and
$\tau_{M-N,M}(x,y,t)$ as \textit{dual} to each other if the solution
$u(-x,-y,-t)$ produced by the function $\tau_{N,M}(-x,-y,-t)$
and the solution generated by $\tau_{M-N,M}(x,y,t)$ are in the same
equivalence class. Note that $\tau_{N,M}(-x,-y,-t)$ is not exactly a
tau-function according to Eq.~\eqref{e:tauphases}, but it can be
transformed to a dual tau-function $\tau_{M-N,M}(x,y,t)$ whose
coefficient matrix~$B$ can be derived from the coefficient matrix~$A$
associated with the tau-function $\tau_{N,M}(x,y,t)$, as we
show next.
                                                                                                                   
Since $A$ is of rank $N$ and in RREF, it can be expressed
as $A=[I_N,G]\,P$, where $I_N$ is the $N \times N$~identify matrix
of pivot columns, $G$ is the $N\times(M-N)$~matrix
of non-pivot columns, and $P$ denotes the $M \times M$ permutation matrix
of $M$~columns of $A$. We augment $A$ with $M-N$ additional rows 
to form the invertible $M \times M$ matrix
\begin{equation}
S = \begin{pmatrix}I_N &G\\
    O &I_{M-N}
  \end{pmatrix}\,P\,,
\label{e:Aedef}
\end{equation}
where $O$ is the $(M-N)\times N$ zero matrix and
$I_{M-N}$ is the $(M-N) \times (M-N)$ identity matrix.
Let $A'$ be the $(M-N)\times M$ matrix obtained by
selecting the last $M-N$ rows of $(S^{-1})^T$. The rank of~$A'$ 
is $M-N$, and the following complementarity relation 
exists between $A$ and~$A'$:
\begin{lemma}
The pivot columns of~$A'$ are labeled by exactly the same
set of indices which label the non-pivot columns of~$A$,
and viceversa.
Moreover, if $A$ is irreducible $A'$~is also irreducible.
\label{A:A'pivot}
\end{lemma}
\begin{proof}
From Eq.~\eqref{e:Aedef} and the fact that
$P^{-1}=P^T$ for a permutation matrix,
we obtain
\begin{equation}
(S^{-1})^T =
  \begin{pmatrix}I_N &O^T\\ -G^T &I_{M-N}
  \end{pmatrix}\,P\,,
\label{e:A'}
\end{equation}
which implies that $A'=[-G^T, I_{M-N}]\,P$.
It is then clear that the pivot columns of $A'P^{-1}$
are its last $M-N$ columns which correspond
to the non-pivot columns of $A\,P^{-1}=[I_N,G]$, and viceversa.
The same correspondence between pivot and non-pivot columns also holds
for~$A$ and~$A'$ because the columns of both matrices are
permuted by the same matrix $P^{-1}$. This proves the first part
of the~Lemma.

To establish that~$A'$ is irreducible,
note first from Definition~\ref{A:D1}
that the permutation of columns preserves irreducibility of a matrix.
Since $A$ is irreducibile, Definition~\ref{A:D1} implies that
all rows or columns of $G$ and~$G^T$ are non-zero.
Therefore the matrix $A'P^{-1}=[-G^T, I_{M-N}]$, and hence~$A'$,
are both irreducible.~%
\end{proof}
Note that although $A'$ is {\em not} in RREF, it be put in RREF by a 
$\mathrm{GL}(N, \Real)$ transformation. Next, we define the matrix~$B$
which is also of rank $M-N$ and irreducible like $A'$, and
whose columns are obtained from~$A'$ as 
\begin{equation}
B[m] = (-1)^mA'[m], \quad m=1,\ldots,M\,.
\label{e:b}
\end{equation}
Then using Eqs. \eqref{e:A'} and \eqref{e:b},
the minors of $A$ can be expressed in terms of the 
complementary minors of~$B$ via (see e.g., Ref.~\cite{gantmacher1959}, p.~21)
\begin{equation}
A(l_1,\dots,l_N)
   = (-1)^\sigma\,\det(P)\,\,B(m_1,\dots,m_{M-N})\,,
\label{e:bminors}
\end{equation}
where $\sigma= M(M+1)/2+N(N+1)/2$, and where the indices
$ m_1\le m_2\le \cdots \le m_{M-N}$ are the complement of
$1\le l_1 \le l_2 \le\cdots l_N$ in $\{1,2,\dots,M\}$.
Furthermore, $B$ plays the role of a coefficient matrix for
the dual tau-function as given by the following lemma.
\begin{lemma}
(Duality)~
If $\tau_{N,M}(x,y,t)$ is the tau-function associated with an irreducible
$N\times M$ coefficient matrix~$A$, then
the matrix~$B$ defined via Eq.~\eqref{e:b} generates a
tau-function~$\tau_{M-N,M}(x,y,t)$ that is dual to~$\tau_{N,M}(x,y,t)$.~%
\label{A:duality}
\end{lemma}
\begin{proof}
Without loss of generality we choose the tau-function~$\tau_{N,M}(x,y,t)$
associated with the given equivalence class of solutions such that
$\theta_{m,0}=0$ for all $m=1,\ldots,M$ in Eq.~\eqref{e:tauphases}.
Then, using Eq.~\eqref{e:bminors} we can express the tau-function as
\begin{subequations}
\begin{equation}
\tau_{N,M}(-x,-y,-t)= (-1)^\sigma\,\det(P)\,
  \e^{-\theta_{1,\dots,M}}\, \tau'_{M-N,M}(x,y,t)\,,
\label{e:tau'}
\end{equation}
where
\begin{equation}
\tau'_{M-N,M}(x,y,t)=  \!\!\!
  \sum\limits_{1\le m_1<m_2<\dots<m_{M-N}\le M} \!\!\!
    V(l_1,\dots,l_N)\,\,
    B(m_1,\dots,m_{M-N})\,\,
    \e^{\theta_{m_1,\dots,m_{M-N}}}\,, 
\label{A:taudet}
\end{equation}
\end{subequations}
with $V(l_1,\dots,l_N)$ denoting the Van~der~Monde determinant
as in Eq.~\eqref{e:tauphases}
and where the sum is now taken over the complementary indices
$m_1,\ldots,m_{M-N}$ instead of $l_1,\ldots,l_N$.
(The number of terms in the sum remains the same
since $\binom{M}{N}=\binom{M}{M-N}$).
It should be clear from Eq.~\eqref{e:u} that
both $\tau_{N,M}(-x,-y,-t)$ and $\tau'_{M-N,N}(x,y,t)$ in Eq.~\eqref{e:tau'}
generate the same solution $u(x,y,t)$ of KPII although
$\tau'_{M-N,M}(x,y,t)$ \textit{is not} a tau-function
as given by Eq.~\eqref{e:tauphases}.
Moreover, all the non-zero minors of~$B$ have the same sign, which is
determined by the sign of~$(-1)^\sigma\det(P)>0$.
Thus, by replacing each Van~der~Monde coefficient $V(l_1,\ldots,l_N)$
by $V(m_1,\ldots,m_{M-N})$ in Eq.~\eqref{A:taudet},
$\tau'_{M-N,M}(x,y,t)$ can be transformed into
a new tau-function~$\tau_{M-N,M}(x,y,t)$ 
associated with the irreducible coefficient matrix~$B$.
Since both $\tau'_{M-N,M}(x,y,t)$ and~$\tau_{M-N,M}(x,y,t)$
are sign-definite sums of the \textit{same} exponential phase 
combinations, they generate solutions that are in the same equivalence 
class. Therefore, the tau-functions $\tau_{N,M}(x,y,t)$ and 
$\tau_{M-N,N}(x,y,t)$ are dual to each other, thus proving the lemma.
\end{proof}
By applying Lemma~\ref{A:duality}, it is easy to show that
part~(i) of Theorem~\ref{A:T1} implies part~(ii) and viceversa.
For example, by applying part~(i) of Theorem~\ref{A:T1} to the
tau-function $\tau_{M-N,M}(x,y,t)$ in Lemma~\ref{A:duality}
one can conclude that as $y \to \infty$, $\tau_{M-N,M}(x,y,t)$ 
generates a solution with exactly $M-N$ line solitons, identified by 
the {\em pivot} indices $g_1,\ldots,g_{M-N}$ of the associated coefficient 
matrix~$B$.\unskip\footnote{Since the ordering of the pivot 
and non-pivot columns of~$B$
is reversed with respect to that of~$A$, if $[i,j]$ with
$i<j$ labels an asymptotic line soliton generated by
$\tau_{M-N,M}(x,y,t)$ as~$y\to\infty$, then $j$ is the pivot index, not~$i$.}
Since $\tau_{M-N,M}(x,y,t)$ is dual to $\tau_{M,N}(x,y,t)$, the solution
generated by $\tau_{M-N,M}(x,y,t)$ is in the same equivalence class
as $u(-x,-y,-t)$. Consequently, the asymptotic line solitons of $u(-x,-y,-t)$
as $y \to \infty$, are labeled by exactly the same indices $g_1,\ldots,g_{M-N}$.
Then it follows that as $y\to-\infty$, there are $M-N$ asymptotic line solitons
of the solution $u(x,y,t)$ generated by $\tau_{N,M}(x,y,t)$. Furthermore,
these line solitons are labeled by the same indices $g_1,\ldots,g_{M-N}$ which
are the non-pivot indices of the coefficient matrix $A$ of the tau-function
$\tau_{N,M}(x,y,t)$,thus proving part~(ii) of Theorem~\ref{A:T1}.
Similarly, one could also prove part~(i) of the Theorem
using part~(ii) and Lemmas~\ref{A:duality}.

Another consequence of Lemma~\ref{A:duality} is that
the dominant pairs of phase combinations for the asymptotic line solitons
of $\tau_{M-N,N}(x,y,t)$ as $y\to\infty$
are the complement of those for the asymptotic line solitons of
the dual tau-function $\tau_{N,M}(x,y,t)$ as $y\to-\infty$.
Thus, if the dominant pair of phase combinations for $\tau_{M-N,M}(x,y,t)$
as $y\to\infty$ along the line~$L_{i,j}$
is given by $\theta_{i,m_2,\dots,m_{M-N}}$ and~$\theta_{j,m_2,\dots,m_{M-N}}$,
the dominant phase combinations
for $\tau_{N,M}(x,y,t)$ as $y\to-\infty$ along~$L_{i,j}$
are $\theta_{i,l_2,\dots,l_N}$ and~$\theta_{j,l_2,\dots,l_N}$,
where the index set $\{l_2,\dots,l_N\}$ is the complement of 
$\{i,j,m_2,\dots,m_{M-N}\}$ in $\{1,\dots,M\}$.

A particularly interesting subclass of $(N_-,N_+)$-soliton solutions 
is obtained by requiring the solutions $u(x,y,t)$ and $u(-x,-y,-t)$ 
to be in the same equivalence class which is generated by ``self-dual'' 
tau-functions. These are the elastic $N$-soliton solutions of KPII,
for which the amplitudes and directions of the $N$ incoming line solitons
coincide with those of the $N$ outgoing line solitons,
as mentioned in section~\ref{s:introduction}.
Thus, the set of incoming line solitons and the set of outgoing line solitons
can both be labeled by the same index pairs $\{[i_n,j_n]\}_{n=1}^N$.
Clearly, in this case we have $N_+=N_-=N$ and $M=2N$.
The detailed properties of the elastic $N$-soliton solution
are studied in Refs.~\cite{GBSCYK,Kodama}.
Here we only mention one result which is a direct consequence of 
Theorem~\ref{A:T1} and the above discussions:
\begin{corollary}
A necessary condition for a set of index pairs
$\{[i_n,j_n]\}_{n=1}^N$ to describe an elastic $N$-soliton solution is
that the indices
$i_1,\dots,i_N$ and $j_1,\dots,j_N$ form a disjoint partition of
the integers $1,\dots,2N$.
\label{C:elastic}
\end{corollary}
\begin{proof}
From part~(i) of Theorem~\ref{A:T1}, the indices $i_1,\dots,i_N$
for the $N$~asymptotic line solitons as $y\to\infty$ label the pivot columns of~$A$,
and from part~(ii) of Theorem~\ref{A:T1}, the indices $j_1,\dots,j_N$
for the $N$~asymptotic line solitons as $y\to-\infty$ label the
non-pivot columns of~$A$.
In order for the $N$~asymptotic line solitons as $y\to-\infty$ to be
the same as those as $y\to\infty$, however, the index pairs $[i_n,j_n]$
must obviously be the same as $y\to\pm\infty$ for all $n=1,\dots,N$.
But the sets of pivot and non-pivot indices of any matrix are of course
disjoint; hence the desired result.
\end{proof}
Note however that the condition in Corollary~\ref{C:elastic} is
necessary but {\em not}
sufficient to describe an elastic $N$-soliton solution. It is indeed
possible to have $N$-soliton solutions where the index pairs
labeling the asymptotic line solitons as $y \to \pm \infty$ form
two different disjoint partition of integers $\{1,2,\ldots,2N\}$.
Such $N$-soliton solutions are not elastic. See, for example the
$2$-soliton solution in Fig.~\ref{f:kps}e.

\catcode`\@ 11
\def\journal#1&#2,#3 (#4){\begingroup \let\journal=\d@mmyjournal {\frenchspacing\sl #1\/\unskip\,} {\bf\ignorespaces #2}\rm, #3 (#4)\endgroup}
\def\d@mmyjournal{\errmessage{Reference foul up: nested \journal macros}}
\def\title#1{{``#1''}}
\def\@biblabel#1{#1.}
\catcode`\@ 12

\end{document}